\documentclass[reprint,superscriptaddress, 
aps,preprintnumbers]{revtex4-1}
\usepackage{bm,bbm}
\usepackage{natbib,textcase}
\usepackage{booktabs}
\usepackage{multirow}
\usepackage{mathtools, amsfonts, amsthm, latexsym, amssymb,dsfont}
\usepackage{braket}
\usepackage[T1]{fontenc}
\usepackage{microtype}
\usepackage[hyperindex,breaklinks]{hyperref}
\usepackage{graphicx}
\usepackage{xcolor}
\usepackage{array}
\usepackage{enumitem}

\usepackage{braket}
\usepackage{tikz}
\usepackage{ascmac}
\usepackage{comment}
\usepackage{stackrel}
\usepackage{accents}
\usepackage{slashed}

\usetikzlibrary{patterns}
\usetikzlibrary{decorations.markings}
\usetikzlibrary{decorations.pathmorphing}
\usetikzlibrary{arrows}

\DeclareMathOperator{\Tr}{Tr}

\usepackage{blkarray}

\usepackage{nicematrix}

\usepackage{tikz}
\usetikzlibrary{intersections, calc,positioning,decorations.pathreplacing,decorations.pathmorphing}
\tikzset{>=latex}

\def\d{{\rm d}}

\def\i{{\rm i}}

\def\CD{{\cal D}}

\def\CN{{\cal N}}

\def\CT{{\cal T}}

\def\BR{\mathbb{R}}

\def\BZ{\mathbb{Z}}

\def\b0{\bm{0}_\perp}

\def\torus{\textbf{T}}

\begin{document}

\title{Entanglement R\'enyi entropy and boson-fermion duality in massless Thirring model}

\author{Harunobu Fujimura}
\author{Tatsuma Nishioka}
\author{Soichiro Shimamori}

\affiliation{
Department of Physics, Osaka University,\\
Machikaneyama-Cho 1-1, Toyonaka 560-0043, Japan
}

\date{\today}

\preprint{OU-HET-1203}

\begin{abstract}
We investigate the second R\'enyi entropy of two intervals in the massless Thirring model describing a self-interacting Dirac fermion in two dimensions.
Boson-fermion duality relating this model to a free compact boson theory enables us to simplify the calculation of the second R\'enyi entropy, reducing it to the evaluation of the partition functions of the bosonic theory on a torus.
We derive exact results on the second R\'enyi entropy, and examine the dependence on the sizes of the intervals and the coupling constant of the model both analytically and numerically.
We also explore the mutual R\'enyi information, a measure quantifying the correlation between the two intervals, and find that it generally increases as the coupling constant of the Thirring model becomes larger.
\end{abstract}

\maketitle

\section{Introduction}\label{sec: Introduction}
Quantum entanglement, a fundamental phenomenon in modern physics characterized by non-local correlations between quantum states separated by distance, has garnered extensive attention not only in quantum information theory but also in high energy physics, condensed matter theory, and gravitation theory.
In quantum field theory (QFT), various measures of quantum entanglement have proven to be indispensable tools for characterizing and classifying  different phases of matter, particularly topological phases \cite{Kitaev:2005dm,Levin:2006zz}, while also capturing universal scaling behaviors in critical systems \cite{Holzhey:1994we,Vidal:2002rm,Calabrese:2004eu,its2005entanglement}.
Moreover, quantum entanglement has found an unexpected connection with gravitational physics via the holographic principle \cite{Ryu:2006bv,Ryu:2006ef}, yielding fresh perspectives on the intricate structures of spacetime, including those governing black hole physics, as well as non-perturbative aspects of QFT.
(See e.g., \cite{Calabrese:2009qy,Casini:2009sr,Nishioka:2009un,Rangamani:2016dms,Nishioka:2018khk} for reviews.)

The entanglement R\'enyi entropy (ERE) is one of the principal measures that quantify the amount of quantum entanglement shared between different parts of a quantum system.
They are a generalization of the more familiar measure known as entanglement entropy, defined by
\begin{align}
    S_n \equiv \frac{1}{1-n}\log \Tr_V \left[ \rho_V^n\right] \ ,
\end{align}
where $\rho_V$ is the reduced density matrix of a subsystem $V$ and $n$ is a parameter that can take on any positive integer. In QFT, the subsystem $V$ is given by a subregion on a time slice, and the reduced density matrix $\rho_V$ can be represented by the Euclidean path integral on a manifold with a cut along $V$.
Hence, the calculation of the ERE amounts to that of the partition function $Z_n$ on the replica manifold, which is an $n$-sheeted manifold branched over $V$.
The codimension-two conical singularity around $\partial V$ often prevents us from performing the path integral analytically on the replica manifold.
The ERE has had exact results for general $n$ only in simple cases to date, such as an interval in two-dimensional conformal field theories (CFTs) \cite{Holzhey:1994we,Calabrese:2004eu}, multiple intervals in free massless fermion theory \cite{Casini:2005rm}, and spherical region in $(d\ge 3)$-dimensional free massless scalar and fermion theories \cite{Klebanov:2011uf}.
Another exactly calculable case is the second ERE ($n=2$) of two intervals in a two-dimensional free compact boson theory \cite{Calabrese:2009ez,Calabrese:2010he,Headrick:2010zt}.
However, when it comes to incorporating interactions, the calculation of the ERE proves to be more challenging, and only a few approximate results have been obtained, such as in the $\text{O}(N)$ models in the large-$N$ limit \cite{Metlitski:2009iyg}, so far \footnote{
By suitably extending the definition, the ERE admits an exact calculation in a certain class of interacting supersymmetric field theories \cite{Nishioka:2013haa}.}.

In this paper, we calculate the ERE exactly in the massless Thirring model describing a self-interacting Dirac fermion in two dimensions \cite{Thirring:1958in};
\begin{align}\label{eq: plane massless Thirring}
    \int \d^2x \, 
    \left[
    \, \i \,\Bar{\psi}\, \slashed{\partial}\, \psi 
    +\frac{\pi }{2}\,\lambda\, (\Bar{\psi}\, \gamma^{\mu}\, \psi)(\Bar{\psi}\, \gamma_{\mu}\, \psi)\,  
    \right]\ . 
\end{align}
This model can be seen as a marginal deformation of a free massless Dirac fermion and enjoys conformal symmetry with central charge $c=1$.
Moreover, it possesses a duality with a free boson theory of a compact radius.
We can leverage the boson-fermion duality to reduce the formidable calculation of  the ERE in the interacting fermionic theory to a more manageable task in the free bosonic theory.
Nevertheless, a subtlety arises when evaluating the ERE with the boson-fermion duality \cite{Headrick:2012fk}.
Since the replica manifold $\Sigma_{n,N}$ associated with the $n$-th ERE of $N$ intervals is a Riemann surface of genus $g=(n-1)(N-1)$, fermionic theories depend on the spin structures along $2g$ one-cycles on $\Sigma_{n,N}$.
On the other hand, the dual bosonic theories can be defined without specifying the spin structures, so their EREs do not appear to match unless $n=1$ or $N=1$.
This apparent paradox is resolved by taking into account the coupling of the dual bosonic theory to a topological field theory (the Kitaev Majorana chain \cite{Kitaev:2000nmw}), and the gauging of a non-anomalous $\BZ_2$ symmetry \cite{Alvarez-Gaume:1986rcs,Alvarez-Gaume:1987wwg,Ginsparg:1988ui,Karch:2019lnn,Tachikawalec}.
We employ this refined perspective on the boson-fermion duality and derive the exact results on the second ERE of two intervals ($n=N=2$) in the massless Thirring model.

This paper is structured as follows.
In section \ref{sec: Review : boson-fermion duality}, we provide a concise review of the modern formulation of boson-fermion duality for bosonic theories with a non-anomalous $\BZ_2$ symmetry in two dimensions.
While the duality extends to higher-genus Riemann surfaces, we concentrate on the torus case.
We express the partition functions of fermionic theories with four distinct spin structures (periodic/anti-periodic boundary conditions along the two one-cycles) in terms of the linear combinations of four bosonic partition functions which incorporate the background $\BZ_2$ gauge fields.
This sets the stage for the application of the Thirring/compact boson duality in the later sections.
Section \ref{sec: Renyi entropy in Thirring model} begins with a brief account of the path integral formulation of the ERE in fermionic theories.
We then proceed to derive exact results on the second ERE of two intervals as well as the mutual R\'enyi information, a measure quantifying the correlation between these intervals.
We also examine their dependence on both interval sizes and the coupling constant analytically in some limits and numerically in broad parameter regions.
Finally, section \ref{ss:Discussion} is dedicated to discussion and potential directions for future works.

\section{Boson-fermion duality}\label{sec: Review : boson-fermion duality}
Boson-fermion duality makes it much easier for us to go back and forth between a bosonic theory and a fermionic one. In particular, we can construct a two-dimensional fermionic theory $\CT_{\text{F}}$ from a bosonic one $\CT_{\text{B}}$ via the fermionization dictionary \cite{Karch:2019lnn,Tachikawalec};
\begin{align}\label{eq: fermionization dictionary}
     \CT_{\text{F}}=\frac{\CT_{\text{B}}\times \mathsf{Kitaev}}{\BZ^\text{B}_2}\ ,
\end{align}
where $\BZ_{2}^{\text{B}}$ is a non-anomalous global symmetry in the bosonic theory. 
In this section, we firstly expand on the procedure incorporating the relation \eqref{eq: fermionization dictionary}.
We next rewrite the partition function of the massless Thirring model in terms of those of a free compact boson model by applying the fermionization dictionary \eqref{eq: fermionization dictionary}.

\subsection{Fermionization on a torus}\label{subsec: fermionization on a trous}
Given a bosonic theory $\CT_\text{B}$ with a $\BZ_2$ symmetry, the construction of a corresponding fermionic theory $\CT_{\text{F}}$ proceeds through the following two steps:
\begin{itemize}
    \item Couple the bosonic theory $\CT_{\text{B}}$ with the spin topological QFT (TQFT) $\mathsf{Kitaev}$.
    \item  
    Gauge the diagonal $\BZ_{2}^{\text{B}}$ symmetry of the theory $\CT_{\text{B}}\times\mathsf{Kitaev}$.
\end{itemize}
The fermionization procedure is applicable on any Riemann surface, but for the sake of simplicity, we will focus our attention on the case of a torus $\torus$ below.

First of all, $\mathsf{Kitaev}$ is the fermionic spin TQFT associated with a given $\BZ_{2}^{\text{B}}$ global symmetry, defined by the partition function
\begin{align}\label{eq: spin TQFT}
    \mathsf{Kitaev}\ : \ (-1)^{\text{Arf}\,[T+\tilde{\varrho}]+\text{Arf}\,[\tilde{\varrho}]}\ , 
\end{align}
where $T\in H^{1}(\torus, \BZ_{2}^{\text{B}})$ is the background gauge field associated with $\BZ_{2}^\text{B}$ symmetry, and characterized by holonomy around the two one-cycles $\gamma_{1}$ and $\gamma_{2}$ of $\torus$;
\begin{align}
    T=(T_{1}, T_{2})\ , \quad T_{I}\equiv\int_{\gamma_I}T\in \{0, 1\}\ , \quad I=1, 2\ . 
\end{align}
Also, $\tilde{\varrho}$ is the spin structure in the fermionic theory $\CT_\text{F}$.
On a torus, there are four choices of the spin structure depending on whether the fermionic field subjects to the periodic (P) or anti-periodic (A) boundary condition for each one-cycle on \torus: $\tilde{\varrho}=\text{PP},\, \text{AP},\, \text{PA},\, \text{AA}$. In addition, Arf$\,[\tilde{\varrho}]$ is the Arf invariant, which takes its value in zero or one depending on the spin structure $\tilde{\varrho}$;
\begin{align}\label{eq: Arf-invariant}
\text{Arf}\,[\tilde{\varrho}]=\left\{
\begin{aligned}
0&\qquad  \tilde{\varrho}= \text{AP}\, ,\, \text{PA}\, ,\, \text{AA} \, , \\
1&\qquad  \tilde{\varrho}=\text{PP}  \, . 
\end{aligned}
\right.
\end{align}
We should note that the insertion of a non-trivial background $\BZ_{2}^{\text{B}}$ gauge field $T$ along a one-cycle $\gamma$ changes the spin structure associated with $\gamma$.
For instance, the holonomy of the $\BZ_{2}^{\text{B}}$ gauge field with its value $(1, 0)$ shifts the spin structures as follows:
\begin{align}
    \text{PP}\to\text{AP}\ , \quad \text{AP}\to\text{PP}\ , \quad \text{PA}\to\text{AA}\ , \quad \text{AA}\to\text{PA} \ , 
\end{align}
and thereby the Arf invariant Arf$\,[(1,0)+\tilde{\varrho}]$ is given by
\begin{align}\label{eq: Arf non-trivial gauge field}
    \text{Arf}\,[(1, 0)+\tilde{\varrho}]=\left\{
\begin{aligned}
0&\qquad  \tilde{\varrho}= \text{PP}\, ,\, \text{AA}\, ,\, \text{PA} \, , \\
1&\qquad  \tilde{\varrho}=\text{AP}  \, . 
\end{aligned}
\right.
\end{align}
Similar considerations apply straightforwardly to the other gauge configurations.
At the level of the partition function, coupling the bosonic theory $\CT_\text{B}$ to the TQFT $\mathsf{Kitaev}$ amounts multiplying the bosonic torus partition function $Z_{\text{B}}[T]$ with the TQFT defined by \eqref{eq: spin TQFT};
\begin{align}
    \CT_{\text{B}}\times \mathsf{Kitaev}\ : \ Z_{\text{B}}[T]\, (-1)^{\text{Arf}\,[T+\tilde{\varrho}]+\text{Arf}\,[\tilde{\varrho}]}\ . 
\end{align}

Next, we proceed to perform the gauging of the $\BZ_{2}^{\text{B}}$ global symmetry.
This means that we promote the background gauge field $T$ to the dynamical one $t$, and sum over all configurations of the gauge field $t$.
This summation is a finite dimensional analog of the usual path integral, and the resulting gauged theory $(\CT_{\text{B}}\times \mathsf{Kitaev})/\BZ_{2}^{\text{B}}$ can be written as follows:
\begin{align}\label{eq: coupled TQFT and summed over gauge field}
    \frac{\CT_{\text{B}}\times \mathsf{Kitaev}}{\BZ^\text{B}_2}\ : \quad \frac{1}{2}\, \sum_{t\in H^{1}(\torus,\, \BZ_{2}^{\text{B}})}Z_\text{B}[t]\, (-1)^{\text{Arf}\,[t+\tilde{\varrho}]+\text{Arf}\,[\tilde{\varrho}]}\ . 
\end{align}
The fermionization dictionary \eqref{eq: fermionization dictionary} asserts that 
\eqref{eq: coupled TQFT and summed over gauge field} is exactly the same as the fermionic torus partition function,
\begin{align}\label{eq: fermionization partition function}
    Z_\text{F}[\tilde{\varrho}]=\frac{1}{2}\, \sum_{t\in H^{1}(\torus,\, \BZ_{2}^{\text{B}})}Z_\text{B}[t]\, (-1)^{\text{Arf}\,[t+\tilde{\varrho}]+\text{Arf}\,[\tilde{\varrho}]}\ . 
\end{align}
For instance, let us consider the case where $\tilde{\varrho} = \text{AA}$. By performing the summation in \eqref{eq: fermionization partition function} with \eqref{eq: Arf-invariant} and \eqref{eq: Arf non-trivial gauge field}, we find
\begin{align}
    Z_\text{F}[\text{AA}]
    &=
    \frac{1}{2}\,(Z_\text{B}[00]+Z_\text{B}[01]+Z_\text{B}[10]-Z_\text{B}[11])\ , \label{eq: fermionization dictionary AA}
\end{align}
where we use a shorthand notation $Z_\text{B}[ab]$ for the partition function $Z_\text{B}[(a,b)]$.

\subsection{Compact boson/massless Thirring duality}
In this subsection, we apply the fermionization dictionary \eqref{eq: fermionization partition function} to the massless Thirring model whose partition function $Z_\text{F}[\tilde{\varrho}, \lambda, \tau]$ on a torus $\torus$ with a modulus $\tau$ is given by 
\begin{align}\label{eq: torus partition function of massless Thirirng}
    Z_\text{F}[\tilde{\varrho}, \lambda, \tau]\equiv\int \CD \bar{\psi}\,\CD \psi\, e^{-I[\tilde{\varrho}, \lambda, \tau]}\ , 
\end{align}
where $I[\tilde{\varrho}, \lambda, \tau]$ is the massless Thirring action defined by \cite{Thirring:1958in}
\begin{align}\label{Thirring action}
    I[\tilde{\varrho}, \lambda, \tau]
    \equiv
    \int_{\torus}\d^2x \, 
    \left[
    \, \i \,\Bar{\psi}\, \slashed{D}_{\tilde{\varrho}}\, \psi 
    +\frac{\pi }{2}\,\lambda\, (\Bar{\psi}\, \gamma^{\mu}\, \psi)(\Bar{\psi}\, \gamma_{\mu}\, \psi)\,  
    \right] \ .   
\end{align}
Here, $\slashed{D}_{\tilde{\varrho}}$ is the Dirac operator with a spin structure $\tilde{\varrho}$ and $\lambda$ is the coupling constant, which will be referred to as the Thirring coupling in this paper. 

The bosonic counterpart is a free compact boson theory whose torus partition function $Z_\text{B}[T, R, \tau]$ is defined by
\begin{align}
    Z_\text{B}[T, R, \tau]\equiv \int \CD \phi\, \exp\left[-\frac{R^{2}}{8\pi}\int_{\torus}\d^2 x\, (D_{T} \phi)^{2} \right]  \ , 
\end{align}
where $\phi$ is the compact boson field with the periodicity $2\pi$, and $T$ is the $\BZ_{2}^{\text{B}}$ gauge field.
Also, $R$ is the compact radius, which is related to the Thirring coupling $\lambda$ as follows \cite{Coleman:1974bu}:
\begin{align}\label{eq: coupling relation}
    \frac{4}{R^2}=1+\lambda \ . 
\end{align}
When the modulus is set to be pure-imaginary, $\tau=\i \,\ell$, the torus partition function $Z_\text{B}[T, R, \i\,\ell]$ is given by the following simple form:
\begin{align}
    Z_\text{B} [00, R, \i\,\ell] 
    &= \frac{1}{ \eta(\i \,\ell )^2 }\  \vartheta_{3}\left(\i\, \frac{2\, \ell}{R^2} \right) \vartheta_{3}\left(\i\, \frac{R^2\,  \ell }{2} \right)\  , \label{eq: boson00} \\
    Z_\text{B} [01, R, \i\,\ell] 
    &= \frac{1}{ \eta(\i \,\ell)^2 }\  \vartheta_{3}\left(\i\, \frac{2\, \ell}{R^2} \right) \vartheta_{2}\left(\i \,\frac{R^2\, \ell }{2} \right) \  , \label{eq: boson01} \\
    Z_\text{B} [10, R, \i\,\ell] 
    &= \frac{1}{ \eta(\i \,\ell )^2 }\  \vartheta_{4}\left(\i\, \frac{2\, \ell}{R^2} \right) \vartheta_{3}\left(\i \,\frac{R^2\, \ell }{2} \right)\ , \label{eq: boson10} \\
    Z_\text{B} [11, R, \i\,\ell] 
    &= \frac{1}{ \eta(\i \,\ell )^2 }\  \vartheta_{4}\left(\i\, \frac{2\, \ell}{R^2} \right) \vartheta_{2}\left(\i \,\frac{R^2\, \ell }{2} \right)\ , \label{eq: boson11}
\end{align}
where $\eta$ is the Dedekind eta function \cite{DiFrancesco:1997nk}
\begin{align}
     \eta(\tau)=q^{\frac{1}{24} }\, \prod_{n=1}\, (1-q^n)\ , \quad q\equiv e^{2\pi \i \,\tau} \ , 
\end{align}
and $\vartheta_{i}$ ($i=2,3,4$) are the Jacobi theta functions \cite{DiFrancesco:1997nk}
\begin{align}
\begin{aligned}\label{eq: Jacobi theta function}
    \vartheta_{2}(\tau)&=\sum_{n \in \BZ +\frac{1}{2}}  q^{\frac{n^2}{2}}  \ , \quad 
    \vartheta_{3}(\tau) =
    \sum_{n \in \BZ }\,  q^{\frac{n^2}{2}} \ ,\\ \quad
    \vartheta_{4}(\tau)&=\sum_{n \in \BZ }\, (-1)^n \,q^{\frac{n^2}{2}}\ .   
\end{aligned}
\end{align}
The fermionization dictionary \eqref{eq: fermionization partition function} allows us to write the fermionic partition function $Z_\text{F}[\tilde{\varrho}, \lambda, \i \ell]$ as the summation of the bosonic ones.
In what follows, we focus on the spin structure $\tilde{\varrho} = \text{AA}$ on the torus, which will be relevant to the calculation of the ERE in section \ref{subsec: analytical calculation}. 
In the case of the compact boson/massless Thirring duality, the formula \eqref{eq: fermionization dictionary AA} becomes;
\begin{align}\label{eq: fermionization partition function Thirring case}
    Z_\text{F}[\text{AA}, \lambda, \i \ell]
    &=
    \frac{1}{2} 
    \left(
    Z_\text{B}[00,R,\i \ell]+Z_\text{B}[01,R,\i \ell] \right.\nonumber\\
    &\hspace{10mm}\left.+
    Z_\text{B}[10,R,\i \ell]-Z_\text{B}[11,R,\i \ell]
    \right)\ ,
\end{align} 
where Thirring coupling $\lambda$ and compact radius $R$ are related as \eqref{eq: coupling relation}. By plugging \eqref{eq: boson00}-\eqref{eq: boson11} into \eqref{eq: fermionization partition function Thirring case} with help of identities in Appendix \ref{sec: Theta function and Eta function identities}, we find
\begin{align}
     Z_\text{F}[\textrm{AA}, \lambda, \i\,\ell]&=
     \frac{1}{\sqrt{2}\, \eta(\i\, \ell )^2} 
     \left[\sum_{j=2}^{4}\,\Xi_{j}(\lambda, \i\,\ell)
     \right]^{\frac{1}{2}} , 
     \label{eq: Z_F_Torus_AA_theta_expression}
\end{align}
where $\Xi_{j}$ ($j=2, 3, 4$) are defined by
\begin{align}
        \Xi_{j}(\lambda, \i\,\ell)\equiv \vartheta_{j}^{2}\left(\i \,\ell\,(1+\lambda)\right)\, \vartheta_{j}^{2}\left(\frac{\i \,\ell}{1+\lambda}\right)\ .
\end{align} 

Before closing this section, we make a comment on the action of the T-duality in the compact boson theory on the massless Thirring model.
We should first note that the reflection positivity restricts the coefficient of the kinetic term in the compact boson theory to be positive, and hence \eqref{eq: coupling relation} leads to a valid range for the Thirring coupling: $\lambda >-1$.
This is however not the end of the story, since there are equivalent points under the T-duality.
Indeed, the T-duality in the compact boson theory maps the radius $R$ to its inverse (see section 3.5 in the literature \cite{Karch:2019lnn} for detail) as
\begin{align}
    \text{T-duality}\quad : \quad R\rightarrow R_{\text{dual}}\equiv \frac{4}{R}\ , 
\end{align}
which acts as the T-duality transformation on the Thirring coupling via \eqref{eq: coupling relation};
\begin{align}
    \text{T-duality}\quad : \quad \lambda\rightarrow \lambda_{\text{dual}}\equiv -\frac{\lambda}{\lambda+1}\ . \label{eq: T_dual_lambda}
\end{align}
It follows that the T-duality swaps the two coupling regions: $(-1, 0]$ and $[0, \infty)$, hence the fundamental domain of the Thirring coupling is either $(-1, 0]$ or $[0, \infty)$.

\section{Entanglement R\'enyi entropy in massless Thirring model}\label{sec: Renyi entropy in Thirring model}
We are now in position to derive the ERE and the mutual R\'enyi information (MRI) in the massless Thirring model.
In section \ref{subsec: Replica trick}, we present a brief review on the replica method for calculating ERE \cite{Casini:2009sr}.
In section \ref{subsec: analytical calculation}, we derive the ERE in the massless Thirring model by utilizing the boson-fermion duality introduced in section \ref{sec: Review : boson-fermion duality}. In section \ref{subsec: plot of renyi entropy}, we examine the ERE and the MRI in the massless Thirring model by varying both the cross-ratio and the Thirring coupling.
In section \ref{subsec: tripartite}, we consider the tripartite information for three intervals, two of which are adjacent to each other.

\subsection{Replica method}\label{subsec: Replica trick}
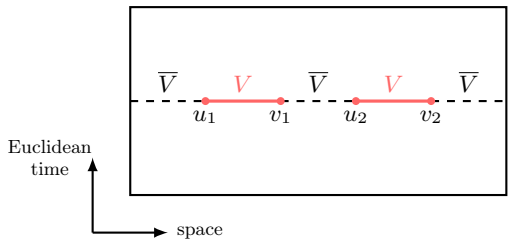
\begin{figure}[t]
  \centering
  \begin{tikzpicture}[scale=0.5]
    \coordinate(u1)at(2,5/2);
    \coordinate(v1)at(4,5/2);
    \coordinate(u2)at(6,5/2);
    \coordinate(v2)at(8,5/2);

    \draw       (1,5/2) node[above] {$\overline{V}$};
    \draw       (5,5/2) node[above] {$\overline{V}$};
    \draw       (9,5/2) node[above] {$\overline{V}$};
    \draw[red!60] (3,5/2) node[above] {$V$};
    \draw[red!60] (7,5/2) node[above] {$V$}; 

    \draw[->,thick](-1,-1) -- (1,-1) node[right]{\scalebox{0.8}{space}};
    \draw[->,thick](-1,-1) -- (-1,1) node[above left=-5mm]{\scalebox{0.8}{\parbox{2cm}{\centering Euclidean\\ time}}};

    \draw[dashed,thick](0,5/2)--(u1) node [below] {$u_{1}$};
    \draw[red!60,very thick](u1)--(v1);
    \draw[dashed,thick](v1)node [below] {$v_{1}$}--(u2) node [below] {$u_{2}$};
    \draw[red!60,very thick](u2)--(v2);
    \draw[dashed,thick](v2)node [below] {$v_{2}$}--(10,5/2);
    \draw[thick](0,0)rectangle(10,5);

    \fill[red!60](u1) circle (3pt) (v1) circle (3pt) (u2) circle (3pt) (v2) circle (3pt);
  \end{tikzpicture}
  \caption{The regions $V$ and $\overline{V}$ in two dimensions for two intervals case. The entangling regions $V$ is shown in red, whose end points are denoted by $u_1$, $v_1$, $u_2$ and $v_2$.}
  \label{fig_region_V}
\end{figure}

To introduce an ERE, we first divide the total space into $V$ and $\overline{V}$. Here, $V$ is the disjoint union of $N$ intervals: 
\begin{align}
    V \equiv [ u_{1}, v_{1} ] \cup [ u_{2}, v_{2}]\cup  \cdots \cup [ u_{N}, v_{N}]\ , 
\end{align}
and $\overline{V}$ is the complementary region to $V$. (See Fig.\,\ref{fig_region_V} for $N=2$ case.) 
Throughout this paper, we assume that the state associated with the total space is the vacuum $\ket{0}$, then the density matrix of the total space is $\rho_{\text{tot}} = \left[ \ket{0}\bra{0}\right]/Z_1$, where $Z_1 \equiv \braket{0|0}$ is the partition function. 
The reduced density matrix of the region $V$ is defined by tracing $\rho_{\text{tot}}$ over all degree of freedom in $\overline{V}$;
\begin{align}
    \rho_{V} \equiv \Tr_{\overline{V}} \left[\rho_{\text{tot}}\right] = \frac{1}{Z_1}\Tr_{\overline{V}} \left[\,\ket{0} \bra{0}\,\right]\ .
\end{align}
The $n$-th ERE $S_{n}(V)$ of the region $V$ is defined by
\begin{align}
    S_{n}(V) \equiv  \frac{1}{1-n} \log  \Tr_{V} \left[\rho^{n}_{V}\right] \  , \quad n \in \BZ_{+}\ .
    \label{eq: Renyi_entropy_definition}
\end{align} 
We should note that, by analytically continuing a positive integer $n$ to a real number and taking the limit $n\to 1$, ERE reduces to the entanglement entropy $S(V)$:
\begin{align}
    \lim_{n \rightarrow 1} S_{n}(V)
    =
    S(V)
    \equiv
    - \Tr_{V} \left[ \rho_{V} \log \rho_{V} \right]
    \ .
\end{align}
We can also introduce the $n$-th MRI $I_{n}(V_{1}, V_{2})$ by
\begin{align}
     I_{n} (V_{1}, V_{2})
    \equiv
    S_{n}(V_{1}) + S_{n}(V_{2}) - S_{n}(V_{1} \cup V_{2})\ , 
    \label{eq: Mutual_Renyi_Info_definition}
\end{align}
where $V_1$ and $V_2$ are disjoint unions of the entangling region $V$ such that $V = V_1 \cup V_2$.
While ERE depends on the choice of the ultraviolet (UV) cutoff in QFT, MRI is known to be finite and independent of the regularization scheme \cite{Headrick:2010zt}.
Intuitively, MRI quantifies how the two regions $V_1$ and $V_2$ are entangled with each other. 

\indent
In order to derive ERE and MRI, we employ the (Euclidean) path integral representation of $\Tr_{V}\left[\rho_{V}^{n} \right]$ \cite{Calabrese:2004eu,Casini:2009sr}:
\begin{align}\label{eq: Replica integral}
    \Tr_{V} \left[\rho^{n}_{V}\right] 
    =
    \frac{Z_{n, N}}{Z_{1}^n}
    \equiv
    \frac{1}{Z_{1}^n}
    \int_{\Sigma_{n,N}}
    \CD \Psi\, \CD \overline{\Psi}\,
    e^{- I[\Psi, \overline{\Psi}]}\ , 
\end{align}
where, $Z_{n, N}$ is the partition function on the $n$-sheeted manifold $\Sigma_{n, N}$ which consists of $n$-copies of the original manifold with an appropriate gluing conditions along with the entangling region $V$ (see Fig. \ref{fig:n_sheeted_manifold}). 
Also, $I[\Psi, \overline{\Psi}]$ is the Euclidean action where $\Psi$ denotes the $n$-vector whose component is the field living on each copy:
\begin{align}
    \Psi\equiv (\psi_1 , \psi_2 , \cdots, \psi_{n})^{\text{T}}\ . 
\end{align}
Furthermore, $\int_{\Sigma_{n,N}}$ means the restricted path integral with the following gluing condition:
\begin{align}\label{eq: gluing condition}
\begin{aligned}
    \Psi(x \in V, t_{\text{E}}=+0) 
    =
    T\  \Psi(x \in V, t_{\text{E}}=-0)\ ,
\end{aligned}
\end{align}
where $T$ is an $n\times n$ matrix and represents the boundary condition of region $V$. 
For a fermionic system, the matrix $T$ is given by \footnote{Here we use the convention such that the cuts are located along the region $V$ in accordance with \cite{Casini:2005rm,Casini:2009sr} (see appendix \ref{appendix:replica method} for details). 
In other literature (e.g., \cite{Headrick:2010zt,Headrick:2012fk,Coser:2015dvp}), however, the cuts are placed along the region $\overline{V}$.
In a fermionic system with the fermion parity symmetry, the two prescriptions are equivalent to each other.
}
\begin{align}\label{eq:T_matrix}
        T
    \equiv
    \begin{pmatrix}
        0          &    1    &                &          \\
                   & \ddots  &\ddots          &          \\
                   &         & \ \ \ \ \ddots &\ \ \ \ 1 \\
        (-1)^{n+1} &         & \ \ \ \        &\ \ \ \ 0 \\
    \end{pmatrix} \, .
\end{align}
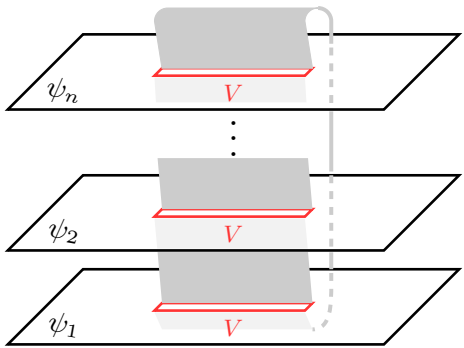
\begin{figure}[t]
  \centering
  \begin{tikzpicture}[scale=0.5]
    \coordinate(left_lower_edge_of_plane)  at (0,0);
    \coordinate(right_lower_edge_of_plane) at (10,0);
    \coordinate(left_upper_edge_of_plane)  at (0+2,2);
    \coordinate(right_upper_edge_of_plane) at (10+2,2);
    \coordinate(a)  at (0,2.5);
    \coordinate(upper_u1) at (4+0.1,1+0.1);
    \coordinate(lower_u1) at (4-0.1,1-0.1);
    \coordinate(upper_v1) at (8+0.1,1+0.1);
    \coordinate(lower_v1) at (8-0.1,1-0.1);

    \draw[line width=1, name path=lower_plane]
    (left_lower_edge_of_plane)--
    (right_lower_edge_of_plane)--
    (right_upper_edge_of_plane)--
    (left_upper_edge_of_plane)--cycle;
    \draw[line width=1, name path=upper_plane]
    ($(left_lower_edge_of_plane)  + (a)$)--
    ($(right_lower_edge_of_plane) + (a)$)--
    ($(right_upper_edge_of_plane) + (a)$)--
    ($(left_upper_edge_of_plane)  + (a)$)--cycle;
    \draw[line width=1, name path=upper_plane]
    ($(left_lower_edge_of_plane)  + 2.5*(a)$)--
    ($(right_lower_edge_of_plane) + 2.5*(a)$)--
    ($(right_upper_edge_of_plane) + 2.5*(a)$)--
    ($(left_upper_edge_of_plane)  + 2.5*(a)$)--cycle;
    
    \draw[line width=1,red!80](lower_u1)--(lower_v1)--(upper_v1)--(upper_u1)--cycle;
    \draw[line width=1,red!80]
    ($(lower_u1) + (a)$)--
    ($(lower_v1) + (a)$)--
    ($(upper_v1) + (a)$)--
    ($(upper_u1) + (a)$)--cycle;
   
    \draw[line width=1,red!80]
    ($(lower_u1) + 2.5*(a)$)--
    ($(lower_v1) + 2.5*(a)$)--
    ($(upper_v1) + 2.5*(a)$)--
    ($(upper_u1) + 2.5*(a)$)--cycle;

    \fill[lightgray!80]
    (upper_u1)--
    (upper_v1)--
    ($(upper_v1)!0.59!($(lower_v1) + (a)$)$)--
    ($(upper_u1)!0.59!($(lower_u1) + (a)$)$)--cycle;
    \fill[lightgray!20]
    ($(upper_v1)!0.63!($(lower_v1) + (a)$)$)--
    ($(upper_u1)!0.63!($(lower_u1) + (a)$)$)--
    ($(lower_u1) + (a) -(0,0.05)$)--
    ($(lower_v1) + (a) -(0,0.05)$)--cycle;
    \fill[lightgray!80]
    ($(upper_u1)+ (a)$)--
    ($(upper_v1)+ (a)$)--
    ($($(upper_v1)+ (a)$)!0.59!($(lower_v1) + 2*(a)$)$)--
    ($($(upper_u1)+ (a)$)!0.59!($(lower_u1) + 2*(a)$)$)--cycle;
    \fill[lightgray!20]
    ($($(upper_v1)+ (a)$)!0.8!($(lower_v1) + 2.5*(a)$)$)--
    ($($(upper_u1)+ (a)$)!0.8!($(lower_u1) + 2.5*(a)$)$)--
    ($(lower_u1) + 2.5*(a) -(0,0.05)$)--
    ($(lower_v1) + 2.5*(a) -(0,0.05)$)--cycle;
    \filldraw[lightgray!80]
    ($(upper_u1) + 2.5*(a)$)--
    ($(upper_v1) + 2.5*(a)$)--
    ($(upper_v1) + 2.5*(a) + 0.5*(a) + (-0.2,0)$) arc (180:90:0.35)--
    ($(upper_u1) + 2.5*(a) + 0.5*(a) + (-0.2,0) + (0.35,0.35)$ ) arc (90:180:0.35)--
    ($(upper_u1) + 2.5*(a) + 0.5*(a) + (-0.2,0)$ )--cycle;
    \draw[line width=1.5,lightgray!80] 
    ($(upper_v1) + 2.5*(a) + 0.5*(a) + (-0.2,0) + (0.35,0.35)$ ) arc (90:0:0.35) [rounded corners]--
    ($(upper_v1) + 2.5*(a) + 0.5*(a) + (-0.2,0) +(0.7,0) + (0, -0.6)$);
    \draw[line width=1.5,dashed,lightgray!80]
    ($(upper_v1) + 2.5*(a) + 0.5*(a) + (-0.2,0) +(0.7,0) + (0, -0.7)$)--
    ($(upper_v1) + 2.5*(a) + 0.5*(a) + (-0.2,0) +(0.7,0) + (0, -2.3)$);
    \draw[line width=1.5,lightgray!80]
    ($(upper_v1) +2.5*(a) + 0.5*(a) + (-0.2,0) +(0.7,0) + (0, -2.4)$)--
    ($(upper_v1) + 2.5*(a) + 0.5*(a) + (-0.2,0) +(0.7,0) + (0, -4)$);
    \draw[line width=1.5,dashed, lightgray!80]
    ($(upper_v1) + 2.5*(a) + 0.5*(a) + (-0.2,0) +(0.7,0) + (0, -4.1)$)--
    ($(upper_v1) +2.5*(a) + 0.5*(a) + (-0.2,0) +(0.7,0) + (0, -6.1)$);
    \draw[line width=1.5,lightgray!80]
    ($(upper_v1) +2.5*(a) + 0.5*(a) + (-0.2,0) +(0.7,0) + (0, -6.15)$)--
    ($(upper_v1) +2.5*(a) + 0.5*(a) + (-0.2,0) +(0.7,0) + (0, -6.6)$);
    \draw[line width=1.5,dashed, lightgray!80]
    ($(upper_v1) + 2.5*(a) + 0.5*(a) + (-0.2,0) +(0.7,0) + (0, -6.7)$)--
    ($(upper_v1) +2.5*(a) + 0.5*(a) + (-0.2,0) +(0.7,0) + (0, -7)$)[rounded corners]
    to [out = -90, in = 0]
    ($(lower_v1)       - 0.2*(a) + (+0.2,0)$);
    \fill[lightgray!20]
    ($(lower_u1)-(0,0.05)$)--
    ($(lower_v1)-(0,0.05)$)--
    ($(lower_v1) - 0.2*(a) + (+0.2,0)$)--
    ($(lower_u1) - 0.2*(a) + (+0.2,0)$)--cycle;

    \draw($($(upper_u1)+ (a)$)!0.5!($(lower_v1)+ 2.5*(a)$)$) node[above=-3mm] {\scalebox{1.5}{$\vdots $}};
    \draw ($(left_lower_edge_of_plane)+(1.5,0.5)$) node {\scalebox{1.2}{\rotatebox{-10}{$\psi_1$}}};
    \draw ($(left_lower_edge_of_plane)+(a)+(1.5,0.5)$) node {\scalebox{1.2}{\rotatebox{-10}{$\psi_2$}}};
    \draw ($(left_lower_edge_of_plane)+2.5*(a)+(1.5,0.5)$) node {\scalebox{1.2}{\rotatebox{-10}{$\psi_n$}}};
    \draw($(upper_u1)!0.5!(lower_v1)$) node[red!80,above=-5mm] {\scalebox{1}{$V$}};
    \draw($($(upper_u1)+ (a)$)!0.5!($(lower_v1)+ (a)$)$) node[red!80,above=-5mm] {\scalebox{1}{$V$}};
    \draw($($(upper_u1)+ 2.5*(a)$)!0.5!($(lower_v1)+ 2.5*(a)$)$) node[red!80,above=-5mm] {\scalebox{1}{$V$}};
  \end{tikzpicture}
  \caption{The sketch of the $n$-sheeted manifold $\Sigma_{n,N}$. Each sheet represents the original spacetime and the sheets are sewed together at the entangling region $V \equiv [ u_{1}, v_{1} ] \cup [ u_{2}, v_{2}]\cup  \cdots \cup [ u_{N}, v_{N}]$ (red).}
  \label{fig:n_sheeted_manifold}
\end{figure}
By substituting \eqref{eq: Replica integral} into the definition of the ERE \eqref{eq: Renyi_entropy_definition}, we can express the $n$-th ERE in terms of the partition function $Z_{n, N}$ on the $n$-sheeted manifold $\Sigma_{n, N}$: 
\begin{align}\label{eq: Replica_trick_formula}
    S_{n}(V) &= \frac{1}{1-n} \log \left[ \frac{Z_{n,N}}{(Z_{1})^n} \right]\ . 
\end{align}

We record a few exact results of EREs derived by the replica method below:
\begin{itemize}
    \item Two-dimensional CFT:    \vspace{2mm} \\
    In the case of one interval, the $n$-th ERE in an arbitrary CFT is given by \cite{Holzhey:1994we}\footnote{Here we drop off the scheme dependent term.}
    \begin{align}\label{eq: existing_result_1_interval_CFT}
        S^{c}_{n}([u,v]) 
        = \frac{c}{6} \left( 1 + \frac{1}{n} \right) \log \left( \frac{v-u}{\epsilon} \right)\  . 
    \end{align}
    where $\epsilon$ is UV cutoff length and $c$ is the central charge in CFT. We emphasize that the coefficient in front of the logarithm depends only on the central charge and the number of sheets, hence is scheme-independent. 
    
    \item Free massless Dirac fermion in two dimensions: \vspace{2mm} \\
    The $n$-th ERE of arbitrary $N$ intervals in the free massless Dirac fermion is calculated in \cite{Casini:2005rm}. In particular, in the case $(n, N)=(2,2)$, the ERE is given by 
    \begin{align}  \label{eq: existing_result_2_interval_free_Dirac}
    \begin{aligned}
        S^{\text{Free}}_{2}(V) 
        &=
        \frac{1}{4} \log \left( \frac{v_{1} - u_{1} }{\epsilon} \cdot \frac{v_{2} - u_{2} }{\epsilon} \right)  + \frac{1}{4} \log (1-x) \, , 
    \end{aligned}
\end{align}
where $\epsilon$ is the UV cutoff length and $x$ is the cross-ratio defined by
    \begin{align}\label{eq: cross-ratio}
    x 
    =
    \frac{(v_{1}-u_{1})(v_{2}-u_{2})}{(u_{2}-u_{1})(v_{2}-v_{1})}\ . 
\end{align}
\end{itemize}
The exact results \eqref{eq: existing_result_1_interval_CFT} and \eqref{eq: existing_result_2_interval_free_Dirac} will be used later.

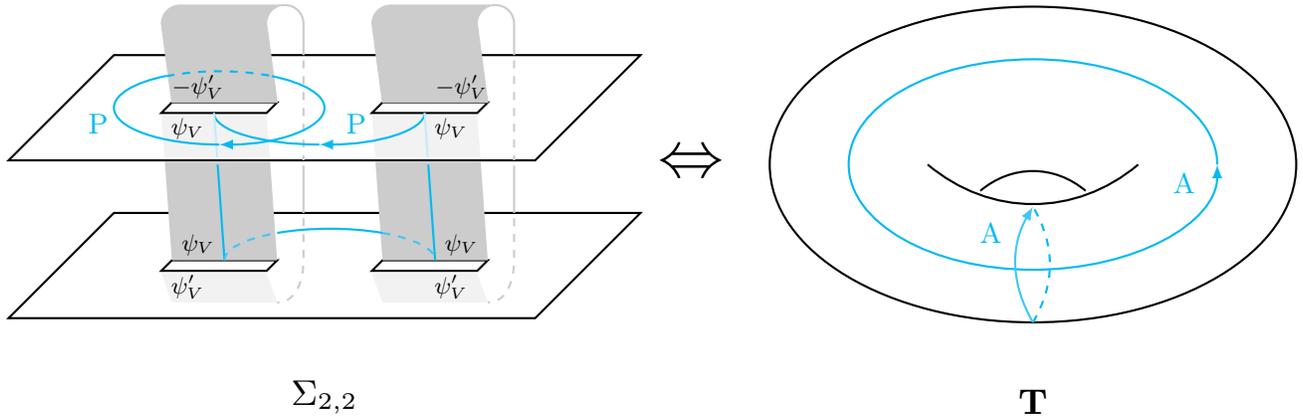
\begin{figure*}[t]
\centering
\begin{tabular}{cc}
  \begin{tikzpicture}[scale=0.7]
    \coordinate(left_lower_edge_of_plane)  at (0,0);
    \coordinate(right_lower_edge_of_plane) at (10,0);
    \coordinate(left_upper_edge_of_plane)  at (0+2,2);
    \coordinate(right_upper_edge_of_plane) at (10+2,2);
    \coordinate(a)  at (0,3);
    \coordinate(upper_u1) at (3+0.1,1+0.1);
    \coordinate(lower_u1) at (3-0.1,1-0.1);
    \coordinate(upper_v1) at (5+0.1,1+0.1);
    \coordinate(lower_v1) at (5-0.1,1-0.1);
    \coordinate(upper_u2) at (7+0.1,1+0.1);
    \coordinate(lower_u2) at (7-0.1,1-0.1);
    \coordinate(upper_v2) at (9+0.1,1+0.1);
    \coordinate(lower_v2) at (9-0.1,1-0.1);
      
    \draw[thick, name path=lower_plane]
    (left_lower_edge_of_plane)--
    (right_lower_edge_of_plane)--
    (right_upper_edge_of_plane)--
    (left_upper_edge_of_plane)--cycle;
    \draw[thick, name path=upper_plane]
    ($(left_lower_edge_of_plane)  + (a)$)--
    ($(right_lower_edge_of_plane) + (a)$)--
    ($(right_upper_edge_of_plane) + (a)$)--
    ($(left_upper_edge_of_plane)  + (a)$)--cycle;
    \draw[thick](lower_u1)--(lower_v1)--(upper_v1)--(upper_u1)--cycle;
    \draw[thick](lower_u2)--(lower_v2)--(upper_v2)--(upper_u2)--cycle;
    \draw[thick]
    ($(lower_u1) + (a)$)--
    ($(lower_v1) + (a)$)--
    ($(upper_v1) + (a)$)--
    ($(upper_u1) + (a)$)--cycle;
    \draw[thick]
    ($(lower_u2) + (a)$)--
    ($(lower_v2) + (a)$)--
    ($(upper_v2) + (a)$)--
    ($(upper_u2) + (a)$)--cycle;

    \fill[lightgray!80]
    (upper_u1)--
    (upper_v1)--
    ($(upper_v1)!0.67!($(lower_v1) + (a)$)$)--
    ($(upper_u1)!0.67!($(lower_u1) + (a)$)$)--cycle;
    \fill[lightgray!20]
    ($(upper_v1)!0.7!($(lower_v1) + (a)$)$)--
    ($(upper_u1)!0.7!($(lower_u1) + (a)$)$)--
    ($(lower_u1) + (a) -(0,0.05)$)--
    ($(lower_v1) + (a) -(0,0.05)$)--cycle;
    \filldraw[lightgray!80]
    ($(upper_u1) + (a)$)--
    ($(upper_v1) + (a)$)--
    ($(upper_v1) + (a) + 0.5*(a) + (-0.2,0)$) arc (180:90:0.35)--
    ($(upper_u1) + (a) + 0.5*(a) + (-0.2,0) + (0.35,0.35)$ ) arc (90:180:0.35)--
    ($(upper_u1) + (a) + 0.5*(a) + (-0.2,0)$ )--cycle;
    \draw[thick,lightgray!80] 
    ($(upper_v1) + (a) + 0.5*(a) + (-0.2,0) + (0.35,0.35)$ ) arc (90:0:0.35) [rounded corners]--
    ($(upper_v1) + (a) + 0.5*(a) + (-0.2,0) +(0.7,0) + (0, -0.6)$);
    \draw[thick,dashed,lightgray!80]
    ($(upper_v1) + (a) + 0.5*(a) + (-0.2,0) +(0.7,0) + (0, -0.7)$)--
    ($(upper_v1) + (a) + 0.5*(a) + (-0.2,0) +(0.7,0) + (0, -2.6)$);
    \draw[thick,lightgray!80]
    ($(upper_v1) + (a) + 0.5*(a) + (-0.2,0) +(0.7,0) + (0, -2.7)$)--
    ($(upper_v1) + (a) + 0.5*(a) + (-0.2,0) +(0.7,0) + (0, -3.6)$);
    \draw[thick,dashed, lightgray!80]
    ($(upper_v1) + (a) + 0.5*(a) + (-0.2,0) +(0.7,0) + (0, -3.7)$)--
    ($(upper_v1) + (a) + 0.5*(a) + (-0.2,0) +(0.7,0) + (0, -4.5)$)[rounded corners]
    to [out = -90, in = 0]
    ($(lower_v1)       - 0.2*(a) + (+0.2,0)$);
    \fill[lightgray!20]
    ($(lower_u1)-(0,0.05)$)--
    ($(lower_v1)-(0,0.05)$)--
    ($(lower_v1) - 0.2*(a) + (+0.2,0)$)--
    ($(lower_u1) - 0.2*(a) + (+0.2,0)$)--cycle;

    \fill[lightgray!80]
    (upper_u2)--
    (upper_v2)--
    ($(upper_v2)!0.67!($(lower_v2) + (a)$)$)--
    ($(upper_u2)!0.67!($(lower_u2) + (a)$)$)--cycle;
    \fill[lightgray!20]
    ($(upper_v2)!0.7!($(lower_v2) + (a)$)$)--
    ($(upper_u2)!0.7!($(lower_u2) + (a)$)$)--
    ($(lower_u2) + (a) -(0,0.05)$)--
    ($(lower_v2) + (a) -(0,0.05)$)--cycle;
    \filldraw[lightgray!80]
    ($(upper_u2) + (a)$)--
    ($(upper_v2) + (a)$)--
    ($(upper_v2) + (a) + 0.5*(a) + (-0.2,0)$) arc (180:90:0.35)--
    ($(upper_u2) + (a) + 0.5*(a) + (-0.2,0) + (0.35,0.35)$ ) arc (90:180:0.35)--
    ($(upper_u2) + (a) + 0.5*(a) + (-0.2,0)$ )--cycle;
    \draw[thick,lightgray!80] 
    ($(upper_v2) + (a) + 0.5*(a) + (-0.2,0) + (0.35,0.35)$ ) arc (90:0:0.35) [rounded corners]--
    ($(upper_v2) + (a) + 0.5*(a) + (-0.2,0) +(0.7,0) + (0, -0.6)$);
    \draw[thick,dashed,lightgray!80]
    ($(upper_v2) + (a) + 0.5*(a) + (-0.2,0) +(0.7,0) + (0, -0.7)$)--
    ($(upper_v2) + (a) + 0.5*(a) + (-0.2,0) +(0.7,0) + (0, -2.6)$);
    \draw[thick,lightgray!80]
    ($(upper_v2) + (a) + 0.5*(a) + (-0.2,0) +(0.7,0) + (0, -2.7)$)--
    ($(upper_v2) + (a) + 0.5*(a) + (-0.2,0) +(0.7,0) + (0, -3.6)$);
    \draw[thick,dashed,lightgray!80]
    ($(upper_v2) + (a) + 0.5*(a) + (-0.2,0) +(0.7,0) + (0, -3.7)$)--
    ($(upper_v2) + (a) + 0.5*(a) + (-0.2,0) +(0.7,0) + (0, -4.5)$)[rounded corners]
    to [out = -90, in = 0]
    ($(lower_v2)       - 0.2*(a) + (+0.2,0)$);
    \fill[lightgray!20]
    ($(lower_u2)-(0,0.05)$)--
    ($(lower_v2)-(0,0.05)$)--
    ($(lower_v2) - 0.2*(a) + (+0.2,0)$)--
    ($(lower_u2) - 0.2*(a) + (+0.2,0)$)--cycle;

    \draw[->,thick,cyan!80]
    ($($(lower_u2) + (a)$)!0.5!($(lower_v2) + (a)$)$) arc[start angle=0, end angle=-90,x radius=2,y radius=0.6];
    \draw[thick,cyan!80]
    ($($($(lower_u2) + (a)$)!0.5!($(lower_v2) + (a)$)$)+(-2,-0.6)$) arc[start angle=-90, end angle=-180,x radius=2,y radius=0.6];
    \draw[thick,cyan!20]
    ($($(lower_u1) + (a)$)!0.5!($(lower_v1) + (a)$)$)--
    ($($($(lower_u1) + (a)$)!0.5!($(lower_v1) + (a)$)$)!0.3!($(upper_u1)!0.5!(upper_v1)$)$);
    \draw[thick,cyan!80]
    ($($($(lower_u1) + (a)$)!0.5!($(lower_v1) + (a)$)$)!0.35!($(upper_u1)!0.5!(upper_v1)$)$)--
    ($(upper_u1)!0.5!(upper_v1)$);
    \draw[dashed,thick,cyan!60]
    ($(upper_u1)!0.5!(upper_v1)$) arc[start angle=180, end angle=120, x radius=2,y radius=0.6];
    \draw[thick,cyan!80]
    (5.1,1.62) arc[start angle=120, end angle=60, x radius=2,y radius=0.6];
    \draw[dashed,thick,cyan!60]
    (7.1,1.62) arc[start angle=60, end angle=0, x radius=2,y radius=0.6];
    \draw[thick,cyan!80]
    ($($($(lower_u2) + (a)$)!0.5!($(lower_v2) + (a)$)$)!0.35!($(upper_u2)!0.5!(upper_v2)$)$)--
    ($(upper_u2)!0.5!(upper_v2)$);
    \draw[thick,cyan!20]
    ($($(lower_u2) + (a)$)!0.5!($(lower_v2) + (a)$)$)--
    ($($($(lower_u2) + (a)$)!0.5!($(lower_v2) + (a)$)$)!0.3!($(upper_u2)!0.5!(upper_v2)$)$);

    \draw[->,thick,cyan!80]
    (4+2,4) arc[start angle=0, end angle=-90, x radius=2, y radius=0.7];
    \draw[thick,cyan!80]
    (4,4-0.7) arc[start angle=-90, end angle=-180, x radius=2, y radius=0.7]
    (4-2,4)   arc[start angle=-180, end angle=-240, x radius=2, y radius=0.7]
    (4+1,4.6)   arc[start angle=-300, end angle=-360, x radius=2, y radius=0.7];
    \draw[dashed,thick,cyan!80]
    (4-1,4.6)   arc[start angle=-240, end angle=-300, x radius=2, y radius=0.7];

    \draw[cyan!80]($($(lower_v1)+ (a)$)!0.5!($(lower_u2)+ (a)$)+(0.7,-0.2)$) node {\scalebox{1.2}{P}};
    \draw[cyan!80](1.7, 3.7) node {\scalebox{1.2}{P}};
    \draw(6,-1.5) node {\scalebox{1.5}{$\Sigma_{2,2}$}};
    \draw($($(upper_u1)+ (a)$)!0.5!($(upper_v1)+ (a)$)+(-0.5,0.3)$) node {$ -\psi'_{V}$};
    \draw($($(lower_u1)+ (a)$)!0.5!($(lower_v1)+ (a)$)+(-0.5,-0.3)$) node {$ \psi_{V}$};
    \draw($(upper_u1)!0.5!(upper_v1)+(-0.5,0.3)$) node {$\psi_{V}$};
    \draw($(lower_u1)!0.5!(lower_v1)+(-0.5,-0.3)$) node {$\psi'_{V}$};

    \draw($($(upper_u2)+ (a)$)!0.5!($(upper_v2)+ (a)$)+(0.5,0.3)$) node {$ -\psi'_{V}$};
    \draw($($(lower_u2)+ (a)$)!0.5!($(lower_v2)+ (a)$)+(0.5,-0.3)$) node {$ \psi_{V}$};
    \draw($(upper_u2)!0.5!(upper_v2)+(0.5,0.3)$) node {$\psi_{V}$};
    \draw($(lower_u2)!0.5!(lower_v2)+(0.5,-0.3)$) node {$\psi'_{V}$};

    \end{tikzpicture}

    \begin{tikzpicture}[scale=0.7]
    \draw(-1.5,5) node {\Huge $\Leftrightarrow$};

    \draw[thick] (5,5)circle[x radius=5,y radius=3];
    \draw[thick] (4,4.5) to [out = 40, in = 140] (6,4.5);
    \draw[thick] (3,5) to [out = -40, in = -140] (7,5);
    
    \draw[<-,thick,cyan!80]
    (5+3.5,5) arc[start angle=0, end angle=-360,x radius=3.5,y radius=2];

    \draw[thick, dashed, cyan!80]
    (5,5-3) to [out = 60, in = -60] (5,4.2);
    \draw[->, thick, cyan!60]
    (5,5-3) to [out = 120, in = -120] (5,4.2);

    \draw[cyan!80](7.5,5) node[below right] {\scalebox{1.2}{A}};
    \draw[cyan!80](4.2, 3.7) node {\scalebox{1.2}{A}};
    \draw(5,0.5) node {\scalebox{1.5}{$\torus$}};
  
  \end{tikzpicture}
  \end{tabular}
  
\caption{$\Tr_{V}\left[ \rho^{2}_{V} \right]$ can be written as the path integral over the two-sheeted manifold $\Sigma_{2,2}$ with the proper gluing condition \eqref{eq: gluing condition} [Left]. The two-sheeted manifold $\Sigma_{2,2}$ is conformally equivalent to the torus $\torus$ [Right]. Fermions are subject to periodic (anti-periodic) boundary conditions along the two cycles in the figure on $\Sigma_{2,2}$ ($\torus$).}
\label{fig_conformal_map_to_torus}
\end{figure*}

Finally, we comment on the spin structure on the two-sheeted manifold $\Sigma_{2,2}$ with two intervals $(n=N=2)$. 
This manifold $\Sigma_{2,2}$ has two non-contractable cycles on which the spin structure $\varrho$ is defined (see the left panel of Fig.\,\ref{fig_conformal_map_to_torus}). Specifying the spin structure $\varrho$ is equivalent to determining whether the fermionic field is subject to the periodic (P) or anti-periodic (A) boundary condition around each cycle. 
Thus, the four spin structures $\varrho=$PP, AP, PA, AA are obtained on $\Sigma_{2,2}$ by assigning one of the two choices along the two cycles. 
Among the four spin structures, however, only $\varrho=\text{PP}$ is compatible with the gluing condition \eqref{eq: gluing condition} \cite{Headrick:2010zt}. 
For $(n, N)=(2,2)$, the gluing condition \eqref{eq: gluing condition} reads 
\begin{align}
\begin{aligned}
    \psi_{1}(x\in V, t_{\text{E}}=+0)&=+\psi_{2}(x\in V, t_{\text{E}}=-0)\ , \\
    \psi_2 (x\in V,t_{\text{E}}=+0) &= -\psi_1 (x\in V,t_{\text{E}}=-0)\ .
\end{aligned}
\end{align}
One can check that the sign of the fermionic field does not change when it moves around each cycle as illustrated on the left side in Fig.\,\ref{fig_conformal_map_to_torus}.
Hence, we will deal with the spin structure $\varrho=\text{PP}$ on the manifold $\Sigma_{2,2}$  in the rest of the paper.

\subsection{Exact results on ERE and MRI}\label{subsec: analytical calculation}
Now, we turn to the derivation of the ERE with $(n, N) = (2, 2)$ in the massless Thirring model \eqref{eq: plane massless Thirring} on the two-dimensional plane $\BR^{2}$.
In general, it is challenging to calculate the ERE analytically in an interacting system by the replica trick as one cannot make the $n$ fields independent by diagonalizing the gluing condition \eqref{eq: gluing condition} by a change of variable.
In what follows, we circumvent this issue by mapping the massless Thirring model \eqref{eq: plane massless Thirring} to a free compact boson by the boson-fermion duality.

It follows from the formula \eqref{eq: Replica_trick_formula} that the ERE with $(n, N) = (2, 2)$ for the massless Thirring model of the coupling $\lambda$ is given by
\begin{align}\label{eq: ERE in massless Thirring model}
    \begin{aligned}
        S_{2}(V, \lambda)
        =
        -\log \left[ \frac{
        Z_{2,2} [\text{PP}, V, \lambda] 
        }{
        \left(Z_{1}\right)^2
        }\right]\ , 
    \end{aligned}
\end{align}
where $Z_1$ and $Z_{2,2}  [\text{PP}, V, \lambda] $ are the partition functions on the two-dimensional plane $\BR^{2}$ and the two-sheeted manifold $\Sigma_{2,2}$ with the spin structure $\varrho=\text{PP}$, respectively.
The partition function $Z_{2,2}$ on $\Sigma_{2,2}$ appears to be difficult to evaluate, but one can map $\Sigma_{2,2}$ to a torus $\torus$ by an appropriate conformal transformation \cite{Dixon:1986qv} and relate $Z_{2,2}$ to the torus partition function $Z_{\text{F}}$ as follows \cite{Lunin:2000yv}:
\begin{align}
\begin{aligned}\label{eq: partition_func_relation_replica_manifold_and_torus}
    \frac{
    Z_{2,2} [\text{PP}, V, \lambda] 
    }{
    (Z_{1})^2
    }
    &=
    \left( 
    \frac{v_{1} - u_{1}}{\epsilon}
    \cdot
    \frac{v_{2} - u_{2}}{\epsilon}
    \right)^{-\frac{1}{4}}\\
    &\hspace{5mm}\cdot
    x^{\frac{1}{4}} \left( 2^8\, x\, (1-x) \right)^{-\frac{1}{12}}
    Z_\text{F}[\text{AA}, \lambda, \i \,\ell]\ , 
\end{aligned}
\end{align}
where $\epsilon$ is the UV cutoff length and $Z_\text{F}[\text{AA}, \lambda, \i \,\ell]$ is the torus partition function of the massless Thirring with the complex structure modulus $\tau = \i \,\ell$ defined in \eqref{eq: torus partition function of massless Thirirng}.
Note that the spin structure $\tilde{\varrho}=\text{AA}$ on $\torus$ is different from the one $\varrho=\text{PP}$ on $\Sigma_{2,2}$ as the periodicity for the fermionic field changes under the conformal transformation \footnote{This is the same as the change of the fermionic boundary condition under the conformal transformation from the two-dimensional plane to the cylinder \cite{Ginsparg:1988ui,Polchinski:1998rr}.}.
Also, $x$ is the cross-ratio defined by \eqref{eq: cross-ratio} and it can be also written in terms of the modulus $\i\,\ell$:
\begin{align}\label{eq: x_l_relation}
    x = \left( \frac{\vartheta_{2}(\i \,\ell)}{\vartheta_{3}(\i \,\ell)} \right) ^4  
    \ , \qquad 
    1-x =  \left( \frac{\vartheta_{4}(\i \,\ell)}{\vartheta_{3}(\i \,\ell)} \right)^4\ ,
\end{align}
where $\vartheta_{j}(\tau)$ $(j = 2, 3, 4)$ are Jacobi theta functions \eqref{eq: Jacobi theta function}.
By substituting \eqref{eq: partition_func_relation_replica_manifold_and_torus} into \eqref{eq: ERE in massless Thirring model}, we can rewrite the ERE with $(n, N)=(2,2)$ in the massless Thirring model as 
\begin{align}
\begin{aligned}\label{eq: ERE_2_2_torus_partition_func_expression}
    S_{2}(V, \lambda) 
    &=
    \frac{1}{4}
    \log
    \left(
    \frac{v_{1} - u_{1}}{\epsilon}
    \cdot
    \frac{v_{2} - u_{2}}{\epsilon}
    \right)\\
    &+
    \frac{1}{12}
    \log
    \left(
    2^8 x^{-2} (1-x)
    \right)
    -
    \log Z_\text{F}[\text{AA}, \lambda, \i \,\ell]\ . 
\end{aligned}
\end{align}
By substituting the boson-fermion duality \eqref{eq: Z_F_Torus_AA_theta_expression} to \eqref{eq: ERE_2_2_torus_partition_func_expression}, we obtain the exact formula of the ERE in the massless Thirring model:
\begin{align}
    S_{2}(V, \lambda) 
    =
    S_{2}(V, 0)
    - \frac{1}{2} \log
    \left[
    \frac{1}{2\vartheta^4_{3}(\i\, \ell)}
    \sum^{4}_{j=2}\,
     \Xi_{j}(\lambda, \i\,\ell)
     \right]\ , \label{eq: ERE_2_2_theta_expression}
\end{align}
where $S_{2}(V, 0)$ is given by
\begin{align}
S_{2}(V, 0)
=
\frac{1}{4} \log \left( \frac{v_{1} - u_{1} }{\epsilon} \cdot \frac{v_{2} - u_{2} }{\epsilon} \right)  + \frac{1}{4} \log (1-x) \ .
\label{eq: ERE_2_2_free_part}
\end{align}
We should note that $S_{2}(V, 0)$ obtained via the boson-fermion duality is consistent with the known result $S_{2}^{\text{Free}}(V)$ in \eqref{eq: existing_result_2_interval_free_Dirac}.
For later convenience, we define the deviation $\Delta S_{2}$ of the ERE from the free part as follows:
\begin{align}
   \Delta S_{2}(x,\lambda)
   \equiv
   S_{2}(V, \lambda) - S_{2}(V, 0)\ . 
\end{align}
Notice that $\Delta S_{2}$ only depends on the cross-ratio $x$ and the Thirring coupling $\lambda$.
Then, we can easily check that $\Delta S_{2}$ is invariant under the modular $S$ transformation: $x \mapsto 1-x$; 
\begin{align}
\Delta S_{2}(x,\lambda)
=
\Delta S_{2}(1-x,\lambda)\ , 
\label{eq: ERE_modular_S_sym}
\end{align}
which ensures that $\Delta S_{2}$ takes the local maximum or minimum at $x=1/2$. Furthermore, the deviation $\Delta S_{2}$ is invariant under the T-duality transformation \eqref{eq: T_dual_lambda}:
\begin{align}\label{eq: duality invariant}
    \Delta S_{2}(x,\lambda)
=
\Delta S_{2}(x,\lambda_{\text{dual}})\ . 
\end{align}
Also, $\Delta S_{2}$ reduces to a simpler form in special cases as shown below:
\begin{itemize}
    \item 
    $\lambda = 1 $ : In this case, we can explicitly write down $\Delta S_{2}(x,\lambda)$ in term of $x$;
        \begin{align}
        \begin{aligned}
            &\Delta S_{2}(x,\lambda = 1)\\
            &=
            \log 2 
            - \log
            \left[
            1+ (1-x)^{\frac{1}{4}} +x^{\frac{1}{4}} - (x(1-x))^{\frac{1}{4}}
            \right]\ . 
        \end{aligned}
        \end{align}
    \item 
    $\lambda \rightarrow -1 + \delta $ : At this point, the system becomes unstable and $\Delta S_{2}(x,\lambda)$ shows a logarithmic divergence; 
        \begin{align}
            \lim_{\delta\rightarrow 0^+}
            \Delta S_{2}(x ,-1+\delta)
            =
            \lim_{\delta\rightarrow 0^+}
            \frac{1}{2} \log \delta 
            =
            - \infty\ .
            \label{eq: ERE_lambda_-1}
        \end{align}
    \item 
    $x \rightarrow 0 $ :
    When the cross-ratio $x$ is close to zero (or, equivalently one), the ERE $S_{2}(V, \lambda)$ approaches the free one $S_{2}(V, 0)$ for any $\lambda$;
    \begin{align}
       \lim_{x \rightarrow  0} \Delta S_{2}(x , \lambda)
       = 0\ .
       \label{eq: ERE_x_0}
    \end{align}
    \item 
    $|\lambda| \ll 1$ : 
    When the coupling $\lambda$ is quite small, we can expand the analytic formula \eqref{eq: ERE_2_2_theta_expression} in powers of $\lambda$.
    After some manipulations, we arrive at the following result;
    \begin{align}\label{eq: ERE_2_2_small_lambda_expression}
    \begin{aligned}
        &\Delta S_{2} (x,\lambda )
        =
        \lambda^2 \,
        \frac{\ell^2}{2 \vartheta^4_{3}(\i \,\ell)} 
        \sum^{4}_{j = 2}\, \vartheta^2_{j} (\i \,\ell)\\
        &\times \left[
            (\partial_{\ell} \vartheta_{j} (\i \,\ell))^2
            -\vartheta_{j}(\i \,\ell)\, \partial^2_{\ell} \vartheta_{j}(\i \,\ell)
            -  \frac{\vartheta_{j}(\i \,\ell) \,\partial_{\ell} \vartheta_{j} (\i \,\ell)}{\ell} 
         \right]\\
        &\hspace{10mm}+
        \mathcal{O} (\lambda^3) \ .
    \end{aligned}
    \end{align}
We should remark that the leading term starts at the second order of $\lambda$.
This can be understood as follows.
In the small coupling region, the duality relation \eqref{eq: T_dual_lambda} is approximated by $\lambda_{\text{dual}} \simeq - \lambda $, and the duality invariance \eqref{eq: duality invariant} implies
\begin{align}
     \Delta S_{2}(x,\lambda)
=
\Delta S_{2}(x,-\lambda)\ , \qquad |\lambda| \ll 1\ ,
\end{align}
which excludes a linear term in $\lambda$ in the expansion.    
\end{itemize}

Finally, we mention the MRI with $(n, N)=(2,2)$ defined by
\begin{align}\label{eq: MRI_2}
    I_{2}(V_{1}, V_{2}, \lambda )
    \equiv
    S_{2}(V_{1}) + S_{2}(V_{2}) - S_{2}(V_{1} \cup V_{2})\ , 
\end{align}
where $V_1$ and $V_2$ are disjoint unions of the entangling region $V$: $V_{1} = [u_{1}, v_{1}]$ and $V_{2} = [u_{2}, v_{2}]$.
The first and second terms are nothing but the EREs with $(n, N)=(2,1)$ given by \eqref{eq: existing_result_1_interval_CFT}. The third term is the ERE with $(n, N)=(2,2)$, and the analytic result has already been obtained in \eqref{eq: ERE_2_2_theta_expression} by the boson-fermion duality.
Therefore, by substituting \eqref{eq: existing_result_1_interval_CFT} and \eqref{eq: ERE_2_2_theta_expression} into \eqref{eq: MRI_2}, we get the MRI with $(n, N)=(2,2)$ in the massless Thirring model;
\begin{align}
    I_{2} (V_{1}, V_{2} ,\lambda)
    =
    I_{2} (V_{1}, V_{2}, 0)
    +
    \frac{1}{2}
    \log
    \left[
    \frac{1}{2\vartheta^4_{3}(\i\, \ell)}
    \sum^{4}_{j=2}\,
      \Xi_{j}(\lambda, \i\,\ell)  
    \right]\ , 
    \label{eq: MRI_2_theta_expression}
\end{align}
where $I_{2} (V_{1}, V_{2}, 0)$ is the contribution from the free theory:
\begin{align}\label{eq: MRI_2_Free}
    I_{2} (V_{1}, V_{2}, 0)
    =
    -\frac{1}{4} \log (1-x)\ .
\end{align}
In particular, when the cross-ratio $x$ is very small and the Thirring coupling $\lambda$ is in the range $1 - \sqrt{3} < \lambda < 1 + \sqrt{3}$, the MRI \eqref{eq: MRI_2_theta_expression} can be approximated as follows;
\begin{align}\label{eq: asymptotic behaviour of MRI}
    I_{2} (x, \lambda)
    \sim
    4 \left(\frac{x}{16} \right)^{\frac{1}{2} \left( 1+\lambda + \frac{1}{1+\lambda} \right)}\ , 
\end{align}
and clearly takes the positive value \footnote{This positivity is quite non-trivial since the MRI does not need to be positive unlike a mutual information.}.
We emphasize that the above result \eqref{eq: asymptotic behaviour of MRI} is perfectly consistent with the universal asymptotic behavior of the MRI \cite[equation (4.26)]{Headrick:2010zt}.

\subsection{Parameter dependence of ERE and MRI}\label{subsec: plot of renyi entropy}
In section \ref{subsec: analytical calculation}, we obtained the exact results of the ERE and MRI with $(n,N) = (2,2)$ in the massless Thirring model.
To grasp the physical meanings of these results, we explore the dependence of the ERE and MRI on the cross-ratio in section \ref{subsubsec: Cross-ratio dependence of ERE} and the Thirring coupling in section \ref{subsubsec: Thirring coupling dependence of ERE and MRI}.

\subsubsection{Cross-ratio dependence} \label{subsubsec: Cross-ratio dependence of ERE}
We begin with examining the cross-ratio dependence of the ERE in \eqref{eq: ERE_2_2_theta_expression} and the MRI in \eqref{eq: MRI_2_theta_expression}.
In Fig.\,\ref{fig_ERE_AA}, we plot the deviation $\Delta S_{2}$ by varying the cross-ratio $x$ while fixing the Thirring coupling $\lambda$ . 
Notice that both the upper and lower panels in Fig.\,\ref{fig_ERE_AA} are symmetric under the map $x\mapsto 1-x$.
Also, $\Delta S_{2}$ takes the maximum or minimum value at $x=1/2$.
These behaviors are in accordance with the modular $S$ symmetry \eqref{eq: ERE_modular_S_sym}.
Furthermore, we can readily check that $\Delta S_{2}$ becomes zero at $x=0$ (or, equivalently $x=1$) for any values of $\lambda$.

\begin{figure}[t]
    \centering
        \includegraphics[width=8cm]{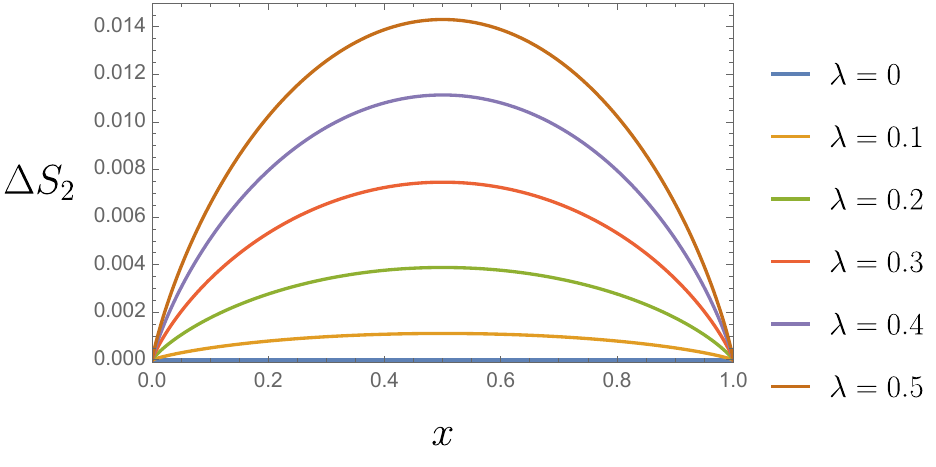}
    \hspace{5mm}
        \includegraphics[width=8cm]{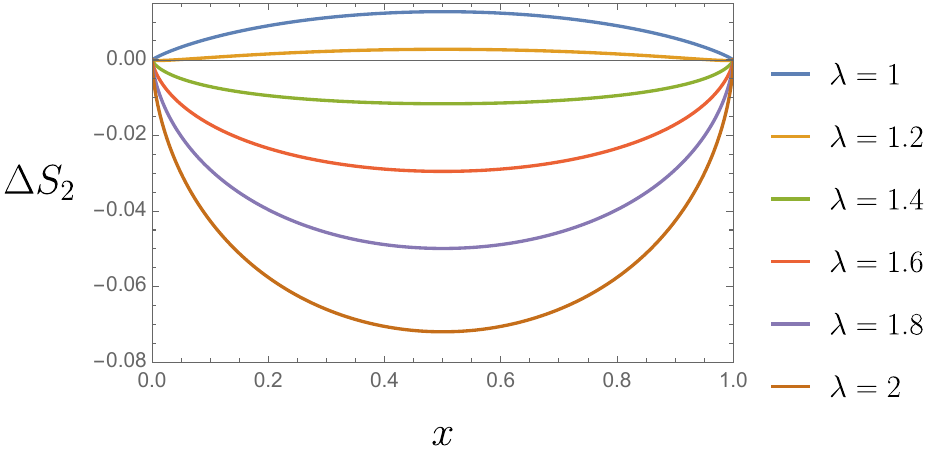}
    \caption{Cross-ratio dependence of the deviation $\Delta S_{2}$ for $\lambda = 0,\ 0.1,\ 0.2,\ 0.3,\ 0.4,\ 0.5$ (upper panel) and $\lambda = 1,\ 1.2,\ 1.4,\ 1.6,\ 1.8,\ 2$ (lower panel).}
    \label{fig_ERE_AA}
\end{figure}
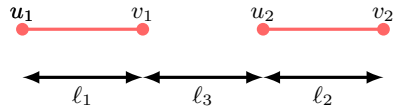
\begin{figure}[t]
  \centering
  \vspace*{1.2cm}
  \begin{tikzpicture}[scale=0.8]
    \coordinate(u1) at (2,0);
    \coordinate(v1) at (4,0);
    \coordinate(u2) at (6,0);
    \coordinate(v2) at (8,0);
    \coordinate(lower_u1) at ($(u1)+(0,-0.8)$);
    \coordinate(lower_v1) at ($(v1)+(0,-0.8)$);
    \coordinate(lower_u2) at ($(u2)+(0,-0.8)$);
    \coordinate(lower_v2) at ($(v2)+(0,-0.8)$);

    \draw       (u1) node[above] {$u_{1}$};
    \draw       (v1) node[above] {$v_{1}$};
    \draw       (u2) node[above] {$u_{2}$};
    \draw       (v2) node[above] {$v_{2}$};
    \draw       (u1) node[above] {$u_{1}$};
    \draw       ($(lower_u1)!0.5!(lower_v1)$) node[below] {$\ell_{1}$};
    \draw       ($(lower_v1)!0.5!(lower_u2)$) node[below] {$\ell_{3}$};
    \draw       ($(lower_u2)!0.5!(lower_v2)$) node[below] {$\ell_{2}$};
    
    \draw[red!60,very thick](u1)--(v1);
    \draw[red!60,very thick](u2)--(v2);
    \draw[<->,very thick](lower_u1)--(lower_v1);
    \draw[<->,very thick](lower_v1)--(lower_u2);
    \draw[<->,very thick](lower_u2)--(lower_v2);
    
    \fill[red!60](u1) circle (3pt) (v1) circle (3pt) (u2) circle (3pt) (v2) circle (3pt);
  \end{tikzpicture}
  \caption{The definitions of the lengths $\ell_1$, $\ell_2$ and $\ell_3$. The red lines are the entangling regions whose end points are denoted by $u_1$, $v_1$, $u_2$ and $v_2$. }
  \label{fig_spacial_relation_between_l1_l2_l3}
\end{figure}
\begin{figure*}[t]
    \begin{minipage}[b]{0.3\linewidth}
        \centering
        \begin{tikzpicture}[scale=0.6]
            \coordinate(u1) at (2,0);
            \coordinate(v1) at (4,0);
            \coordinate(u2) at (6,0);
            \coordinate(v2) at (9,0);
        
            \draw       ($($(u1)!0.5!(v1)$)+(0,0.5)$) node {$\ell_{1}$};
            \draw       ($($(u2)!0.5!(v2)$)+(0,0.5)$) node {$\ell_{2}$};
        
            \draw       ($(v1)!0.5!(u2)$)  node {\rotatebox{90}{\large $\approx$}};
            \draw       ($($(u1)!0.5!(v2)$)+(0,-1)$)  node {$x=0$};
            \draw       ($($(u1)!0.5!(v2)$)+(0,-2)$)  node {($\ell_{3}=\infty$)};
            \draw[red!60,very thick](u1)--(v1);
            \draw[red!60,very thick](u2)--(v2);
            
            \fill[red!60](u1) circle (3pt) (v1) circle (3pt) (u2) circle (3pt) (v2) circle (3pt);
      \end{tikzpicture}
    \end{minipage}
    \hspace{0.03\linewidth}
    \begin{minipage}[b]{0.3\linewidth}
        \centering
        \begin{tikzpicture}[scale=0.6]
            \coordinate(u1) at (2,0);
            \coordinate(v1) at (4,0);
            \coordinate(u2) at ($(4,0) + ({2*sqrt(2)-2},0)$);
            \coordinate(v2) at ($(7,0) + ({2*sqrt(2)-2},0)$);
        
            \draw       ($($(u1)!0.5!(v1)$)+(0,0.5)$) node {$\ell_{1}$};
            \draw       ($($(u2)!0.5!(v2)$)+(0,0.5)$) node {$\ell_{2}$};
            \draw       ($($(u1)!0.5!(v2)$)+(0,-1)$)  node {$x=\frac{1}{2}$};
            \draw       ($($(u1)!0.5!(v2)$)+(0,-2.3)$)  node { };
            
            \draw[red!60,very thick](u1)--(v1);
            \draw[red!60,very thick](u2)--(v2);
            
            \fill[red!60](u1) circle (3pt) (v1) circle (3pt) (u2) circle (3pt) (v2) circle (3pt);
        \end{tikzpicture}
    \end{minipage}
    \hspace{0.03\linewidth}
    \begin{minipage}[b]{0.3\linewidth}
        \centering
        \begin{tikzpicture}[scale=0.6]
            \coordinate(u1) at (2,0);
            \coordinate(v1) at (4,0);
            \coordinate(v2) at (7,0);

            \draw       ($($(u1)!0.5!(v2)$)+(0,0.5)$) node {$\ell_{1} + \ell_{2}$};
            \draw       ($($(u1)!0.5!(v2)$)+(0,-1)$)  node {$x=1$};
            \draw       ($($(u1)!0.5!(v2)$)+(0,-2)$)  node {($\ell_{3}=0$)};
            \draw[red!60,very thick](u1)--(v1);
            \draw[red!60,very thick](v1)--(v2);
            
            \fill[red!60](u1) circle (3pt) (v2) circle (3pt);
        \end{tikzpicture}
    \end{minipage}
    \caption{Locations of the two intervals in the case of $x=0$ [Left], $x=1/2$ [Middle] and $x=1$ [Right]. In this figure, we fix the interval lengths $\ell_1$ and $\ell_2$, hence the limits $x\to 0$ and $x\to 1$ are equivalent to the ones $\ell_3\to \infty$ and $\ell_3\to 0$, respectively. When $x=0$, the two intervals are decoupled with each other, and the ERE can be written as the sum of the ones on the single interval \eqref{eq: l3=infty}. The contribution of the Thirring interaction to the ERE is locally maximized (or minimized) at $x=1/2$. When $x=1$, the two intervals are merged into the single interval with the length $\ell_1 +\ell_2$, and the ERE is given by \eqref{eq: ERE_l3->0}.
    }
    \label{fig_region_V_spacial_relation_l3}
\end{figure*}
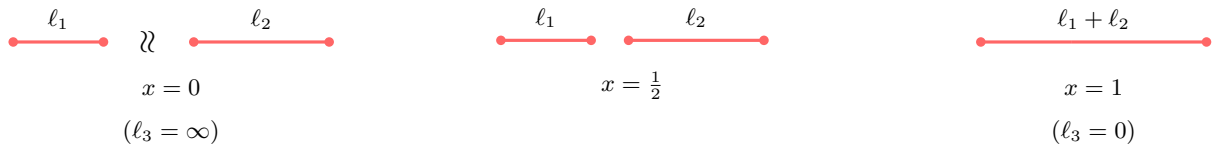

To elucidate the physical meaning of this phenomenon, we rewrite the cross-ratio $x$ defined by \eqref{eq: cross-ratio} as follows;
\begin{align} \label{eq: x_l1_l2_l3_expression}
 x=
    \frac{\ell_{1} \ell_{2}}{(\ell_{1} + \ell_{3})(\ell_{2}+\ell_{3})}\ , 
\end{align}
where $\ell_{1}, \ell_{2}$ and $\ell_3$ are the lengths of the interval defined by (see also Fig. \ref{fig_spacial_relation_between_l1_l2_l3})
\begin{align}
    \ell_{1}\equiv v_{1} - u_{1}\quad , \quad \ell_{2}\equiv  v_{2} - u_{2}      \quad  , \quad \ell_{3}\equiv  u_{2} - v_{1}       \ .
\end{align}
From this expression, we can easily see that the limit $x \rightarrow 0$ is equivalent to the following limits of the interval lengths;
\begin{align}\label{eq: x->0}
    x\rightarrow 0 \qquad \Longleftrightarrow \qquad \frac{\ell_{1}}{\ell_{3}}\rightarrow 0 \quad \text{or} \quad \frac{\ell_{2}}{\ell_{3}}\rightarrow 0\ .  
\end{align}
In the following, we concentrate on the two kinds of limits: $\ell_{1}\rightarrow 0$ and $\ell_{3}\rightarrow \infty$.
\begin{itemize}
    \item $\ell_{1}\rightarrow 0$:\vspace{1mm} \\
    In the limit $\ell_{1}\rightarrow 0$, the entangling region $V$ becomes the single interval $[u_2 , v_2]$, and the ERE with  $(n, N)=(2,2)$ in the massless Thirring model becomes the ERE with $(n, N)=(2,1)$ in the free theory,\footnote{Recall that the Thirring interaction does not contribute to the central charge.}
        \begin{align}\label{eq: l1=0}
        S_{2} (V, \lambda ) \ \longrightarrow \ S^{c=1}_{2} ([u_{2},v_{2}])
        = \frac{1}{4} \log \left( \frac{\ell_2}{\epsilon} \right)  , 
        \end{align}
    hence the deviation $\Delta S_{2}$ must vanish under the limit $\ell_{1}\rightarrow 0$ by definition. 
    \item  $\ell_{3}\rightarrow \infty$:\vspace{1mm}\\ 
    In the limit $\ell_{3}\rightarrow \infty$, the two intervals are decoupled to each other, and the correlation between these intervals tends to vanish. Therefore, the ERE with $(n, N)=(2,2)$ is reduced to the sum of the EREs with $(n, N)=(2,1)$ in the free theory as follows;
    \begin{align}\label{eq: l3=infty}
        S_{2} (V, \lambda )\  \longrightarrow \ S^{c=1}_{2} ([u_{1},v_{1}])+S^{c=1}_{2} ([u_{2},v_{2}]) . 
    \end{align}
    That explains why $\Delta S_{2}$ goes to zero in the limit $\ell_{3} \rightarrow \infty$.
\end{itemize}
Here, it is instructive to compare the ERE in the limit $x\to 1$ with \eqref{eq: l3=infty}.
In a similar manner to $x\to 0$, the limit $x\to1$ can be rephrased as
\begin{align}\label{eq: x->1}
    x\rightarrow 1 \qquad \Longleftrightarrow \qquad \frac{\ell_{3}}{\ell_{1}}\rightarrow 0 \quad \text{and} \quad \frac{\ell_{3}}{\ell_{2}}\rightarrow 0 \ .  
\end{align}
We should notice that we can realize the limit $x\to1$ by putting the two intervals together: $\ell_{3}\to 0 $ with the interval lengths $\ell_1$ and $\ell_2$ fixed.
In this case, the ERE with $(n, N)=(2,2)$ in the massless Thirring model reduces to the  ERE with $(n, N)=(2,1)$ of the interval length $\ell_1 +\ell_2$;
\begin{align}\label{eq: ERE_l3->0}
         S_{2} (V, \lambda )\  \longrightarrow \ S^{c=1}_{2} ([u_{1},v_{2}])
        = \frac{1}{4} \log \left( \frac{\ell_{1}+\ell_{2}}{\epsilon} \right)\  .
\end{align}
Figure \ref{fig_region_V_spacial_relation_l3} illustrates the behaviors of the ERE in varying the cross-ratio $x$ (or the distance $\ell_{3}$ between two intervals) while keeping the interval lengths $\ell_{1}$ and $\ell_{2}$.

\begin{figure}[t]
    \centering
    \includegraphics[width=8cm]{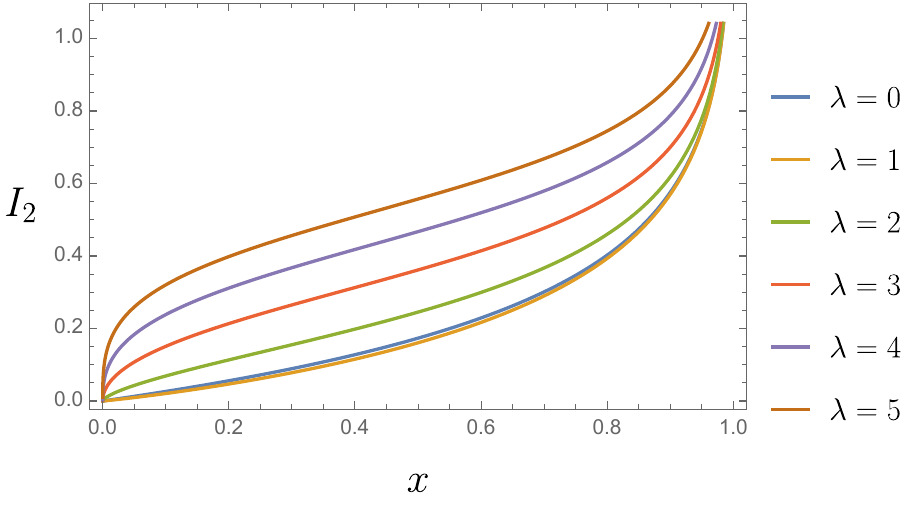}
    \caption{Cross-ratio dependence of the MRI for $\lambda=0,\, 1,\, 2,\, 3,\, 4,\, 5$.}
    \label{fig_MRI_x}
\end{figure}

Finally, we plot the MRI with $(n, N)=(2,2)$ in \eqref{eq: MRI_2_theta_expression} by changing the cross-ratio $x$ in Fig.\,\ref{fig_MRI_x}. In particular, from \eqref{eq: l3=infty} and \eqref{eq: ERE_l3->0}, the MRI with $(n, N)=(2,2)$ behaves under the limits $\ell_3 \to \infty$ and $\ell_3\to 0$ as follows;
\begin{align}
    \begin{aligned}
             I_{2} (V_{1}, V_{2} ,\lambda)\  &\longrightarrow\ 0 , &\text{as}\ \ \ell_{3}\to \infty\ ,  \\
         I_{2} (V_{1}, V_{2} ,\lambda)\  &\longrightarrow\ \frac{1}{4}\log\left[\frac{\ell_{1}\ell_{2}}{\epsilon (\ell_{1}+\ell_2)}\right]=\infty , &\text{as}\ \ \ell_{3}\to 0\ ,
    \end{aligned}
\end{align}
which are completely consistent with the plot in Fig. \ref{fig_MRI_x}.

\subsubsection{Coupling dependence}\label{subsubsec: Thirring coupling dependence of ERE and MRI}
\begin{figure*}[t]
\vspace*{1.2cm}
        \centering
        \includegraphics[height=5cm]{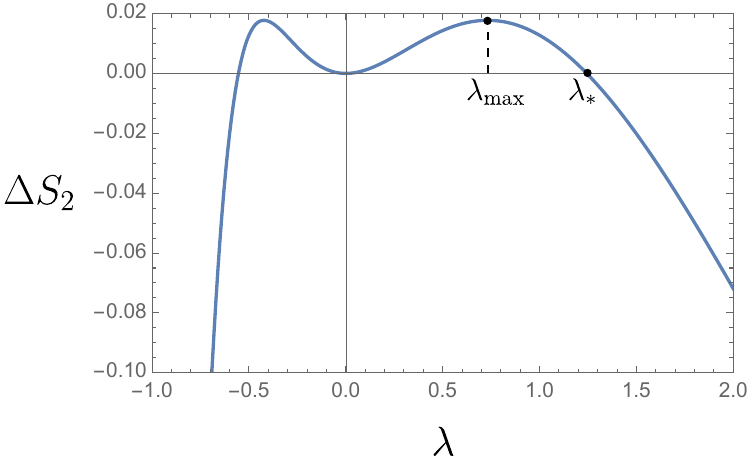}
         \hspace{1cm}
        \centering
        \includegraphics[height=5cm]{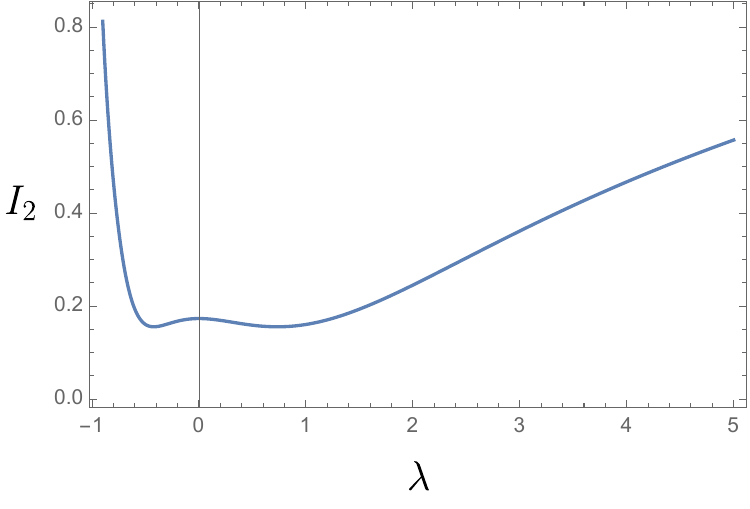}
    \caption{The Thirring coupling dependence of the deviation $\Delta S_{2} (x=1/2,\lambda )$ is shown in the left panel and the MRI $I_{2} (x=1/2, \lambda)$ in the right panel.}
    \label{fig_ERE_MRI_x=1:2}
\end{figure*}
We move on to study the Thirring coupling dependence of the ERE in \eqref{eq: ERE_2_2_theta_expression} and the MRI in \eqref{eq: MRI_2_theta_expression} with keeping the cross-ratio $x=1/2$.
In Fig. \ref{fig_ERE_MRI_x=1:2}, we plot $\Delta S_{2}(x=1/2,\lambda)$ by changing $\lambda$.
Notice that the positive coupling $\lambda\geq0$ can be identified with the negative one $\lambda_{\text{dual}}=-\lambda/(1+\lambda)\leq0$ via the T-duality \eqref{eq: T_dual_lambda}.
Therefore, it is sufficient to consider only the positive coupling region $\lambda\geq0$.
Let $\lambda_{\text{max}}>0$ be the Thirring coupling that maximizes $\Delta S_{2}(x=1/2, \lambda)$.
Then, we observe that $\Delta S_{2} (x=1/2,\lambda )$ monotonically increases in the range $0\leq \lambda \leq \lambda_{\text{max}}$, and decreased in $\lambda\geq \lambda_{\text{max}}$. 
Also, we find out that there is a special point $\lambda_{*}\cong1.244$ where $\Delta S_{2} (x=1/2,\lambda )=0$, namely the ERE in the massless Thirring model is equal to the one in the free theory. Such a special point $\lambda_{*}$ also appear for another value of cross-ratio $x$. Finally, we also plot the Thirring coupling dependence of the MRI in Fig. \ref{fig_ERE_MRI_x=1:2}. 
The MRI in the massless Thirring model is always positive for any value of $\lambda$ and shows a logarithmic divergence at $\lambda = \infty$.

\subsection{Tripartite information}\label{subsec: tripartite}
Let the entangling region $V$ consist of three intervals $A$, $B$ and $C$.
Then, the $n$-th tripartite information $I_{n}(A,B,C)$ is defined by 
\begin{align}
    I_{n}(A,B,C)
    &\equiv
    I_n (A,B) + I_n (A,C) - I_n(A,B\cup C)\ .\label{eq:tripartite_definition}
\end{align}
In topological field theory, the tripartite information $I_1(A,B,C)$ has been studied as topological entanglement entropy \cite{Kitaev:2005dm}.
In holographic theory, $I_1(A,B,C)$ is shown to be negative \cite{Hayden:2011ag, Headrick:2013zda}.
In contrast to $I_1(A,B,C)$, the $n$-th tripartite information for $n\neq 1$ has, however, not been well-investigated.
In what follows, we consider the $n=2$ case and examine the coupling dependence of the second tripartite information $I_2(A,B,C)$ in the massless Thirring model.

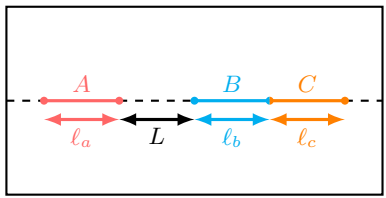
\begin{figure}[t]
  \centering
  \begin{tikzpicture}[scale=0.5]
    \coordinate(ua)at(1,5/2);
    \coordinate(va)at(3,5/2);
    \coordinate(ub)at(5,5/2);
    \coordinate(vb)at(7,5/2);
    \coordinate(vc)at(9,5/2);
    \coordinate(epsilon)at(0,3pt);
    \coordinate(a)at(0,0.5);

    \draw[dashed,thick](0,5/2)--(ua);
    \draw[red!60,very thick](ua)--(va);
    \draw[dashed,thick](va)--(ub);
    \draw[<->,very thick, red!60]($(ua)-(a)$)--($(va)-(a)$);
    \draw[<->,very thick]($(va)-(a)$)--($(ub)-(a)$);
    \draw[<->,very thick, cyan]($(ub)-(a)$)--($(vb)-(a)$);
    \draw[<->,very thick, orange]($(vb)-(a)$)--($(vc)-(a)$);
    
    \draw[cyan,very thick](ub)--(vb);
    \draw[orange,very thick](vb)--(vc);
    \draw[dashed,thick](vc)--(10,5/2);
    \draw[thick](0,0)rectangle(10,5);

    \fill[red!60](ua) circle (3pt) (va) circle (3pt);
    \fill[cyan](ub) circle (3pt);
    \fill[orange] (vc) circle (3pt);
    \fill[cyan] ($(vb)-(epsilon)$)--($(vb)+(epsilon)$) arc [radius=3pt, start angle = 90, end angle=270]--cycle;
    \fill[orange] ($(vb)-(epsilon)$)--($(vb)+(epsilon)$) arc [radius=3pt, start angle = 90, end angle=-90]--cycle;

    \node[red!60] at ($(ua)!0.5!(va)$) [above] {$A$};
    \node[cyan] at ($(ub)!0.5!(vb)$) [above] {$B$};
    \node[orange] at ($(vb)!0.5!(vc)$) [above] {$C$};
    \node[red!60] at ($(ua)!0.5!(va)-(a)$) [below] {$\ell_a$};
    \node[cyan] at ($(ub)!0.5!(vb)-(a)$) [below] {$\ell_b$};
    \node[orange] at ($(vb)!0.5!(vc)-(a)$) [below] {$\ell_c$};
    \node at ($(va)!0.5!(ub)-(a)$) [below] {$L$};
  \end{tikzpicture}
  \caption{The tripartite information for the three intervals $A$, $B$, and $C$ with $B$ and $C$ adjacent. We denote the lengths of the intervals $A$, $B$, $C$, and the distance between $A$ and $B$ as $\ell_a$, $\ell_b$, $\ell_c$ and $L$ respectively.}
  \label{fig:V_tripartite}
\end{figure}

We focus on the case where the two intervals $B$ and $C$ are adjacent, namely, the region $B\cup C$ is just a single interval (see Fig.\,\ref{fig:V_tripartite}).
In this case, the second tripartite information $I_{2}(A,B,C)$ becomes a linear combination of the three mutual R\'enyi informations. 

\begin{figure}[t]
    \centering
    \includegraphics[width=9cm]{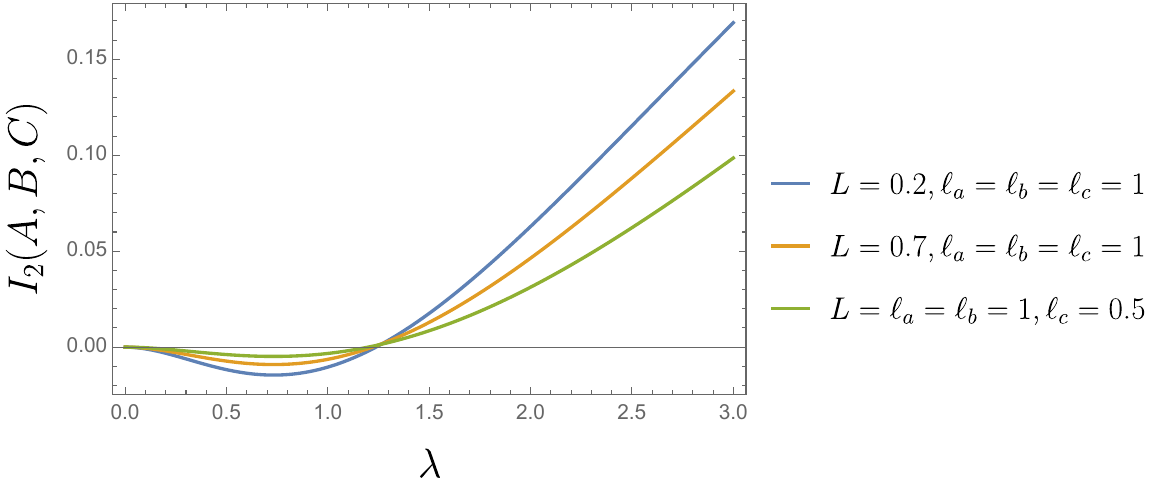}
    \caption{The Thirring coupling dependence of the second tripartite information $I_2(A,B,C)$. We plot the three different configurations of the regions $A$, $B$ and $C$.}
    \label{fig:Tripartite_lambda}
\end{figure}

For the massless Thirring model, we have already derived the mutual R\'enyi information \eqref{eq: MRI_2_theta_expression}, thus we can calculate the second tripartite information $I_2(A,B,C)$ for $B$ and $C$ being adjacent.
In Fig.\,\ref{fig:Tripartite_lambda}, we plot the $\lambda$ dependence of the second tripartite information for several configurations of the three intervals.
We find that the second tripartite information vanishes at $\lambda=0$, $I_2(A,B,C)=0$, which is consistent with the extensive properties of ERE for the free massless fermion \cite{Casini:2008wt}.
We also note that the the sign of the second tripartite information is indefinite, i.e., it takes negative value for small $\lambda$ while it becomes positive for large $\lambda$. It may be intriguing to compare our result with the one obtained by the holographic formula \cite{Dong:2016fnf}.

\section{Conclusion and Discussion}\label{ss:Discussion}
In this paper, we derived the exact results on the ERE and MRI with $(n, N) = (2,2)$ in the massless Thirring model by using the replica method and the boson-fermion duality.
We first rewrote the ERE \eqref{eq: ERE in massless Thirring model} with $(n, N) = (2,2)$ in terms of the torus partition function of the massless Thirring model via the conformal map \eqref{eq: partition_func_relation_replica_manifold_and_torus}.
Then, we exploited the boson-fermion duality between the massless Thirring model and a free compact boson theory, and expressed the torus partition function of the former in terms of those of the latter as in \eqref{eq: fermionization partition function Thirring case}.
We also explored the physical properties of the ERE and MRI with $(n, N) = (2,2)$ in the massless Thirring model by examining the dependence on the cross-ratio and Thirring coupling in section \ref{subsec: analytical calculation} and \ref{subsec: plot of renyi entropy}.
The ERE grows as $\lambda$ increases in the small $\lambda$ regions while it takes the maximum value at the critical point $\lambda = \lambda_\text{max}$ and starts to decrease monotonically in the large $\lambda$ region.
It would be interesting to check if this result can be reproduced in a perturbative calculation with respect to the Thirring coupling $\lambda$. 
On the other hand, we observe that the MRI is non-negative and increases monotonically in the large $\lambda$ region.
Since the MRI quantifies the correlation between two subsystems, this behavior is accordance with our intuition such that quantum entanglement between the two intervals becomes large as the coupling $\lambda$ increases.

This work opens several interesting directions for further studies.
While we only dealt with the massless Thirring model, this serves as the first step toward applying the boson-fermion duality to examine the ERE and MRI.
We believe that the boson-fermion duality will become a fundamental tool for comprehending quantum information quantities and will contribute to a deeper understanding of the dynamics of QFTs.

We can generalize our analysis to include multiple replica sheets or intervals.
As mentioned in Introduction, the $n$-sheeted manifold $\Sigma_{n,N}$ transforms into a Riemann surface of genus $g=(n-1)(N-1)$ through a conformal map, and the boson-fermion duality also applies to such cases \cite{Karch:2019lnn}. Since the partition function of a free compact boson on a higher genus Riemann surface has been previously studied in \cite{Alvarez-Gaume:1986rcs,Alvarez-Gaume:1987wwg,Dijkgraaf:1987vp}, it is possible to extend our work to higher genus cases given a conformal map from a singular Riemann surface $\Sigma_{n,N}$ to a smooth one as in \cite{Lunin:2000yv} for the $n=N=2$ case. 

Furthermore, our work may be extended to the Thirring model with a mass term.
Under the boson-fermion duality, the fermion mass term is known to be mapped to the sine-Gordon potential \cite{Coleman:1974bu}.
When the mass term is small, we can take into account the mass-correction to the ERE and MRI as a relevant perturbation of CFT. (For the free massive fermion theory, a perturbative calculation has been performed in \cite{Casini:2005rm, Herzog:2013py}.)
It is intriguing to apply the conformal perturbation theory to the massive Thirring model and derive the corrections to the ERE and MRI obtained in this paper. 

In this paper, we delved into the specific spin structure $\rho= \text{PP}$ on the two-sheeted manifold $\Sigma_{2,2}$ since it arises in the $(n, N)=(2,2)$ ERE due to the gluing condition \eqref{eq: gluing condition} in the replica method.
Changing the spin structure $\varrho=$ PP to another may be achieved by the insertion of the $\BZ_{2}$ fermion parity operator $(-1)^{F_C}$ which acts on a fermionic field in a subregion $C$ as $(-1)^{F_C}: \psi(x) \to -\psi(x)$ for $x\in C$, and quantum information measures associated with the other spin structures $\varrho=$ AA, AP, PA on $\Sigma_{2,2}$ are discussed in \cite{Fagotti:2010yr, Coser:2015dvp}.
Our method can also be adapted to the calculations of such measures without difficulty.

In the moduli space of $c=1$ bosonic theory, there are some special points where supersymmetries become enhanced. On the one hand, the $n$-th supersymmetric R\'enyi entropy \cite{Nishioka:2013haa} of arbitrary intervals in the $\CN=(2,2)$ superconformal field theory has been derived in \cite{Giveon:2015cgs,Mori:2015bro}.
It may be interesting to compare the result obtained in this paper with the supersymmetric R\'enyi entropy.

\acknowledgments
We are grateful to an anonymous referee of Physical Review D for suggesting us to consider the tripartite information. The work of T.\,N. was supported in part by the JSPS Grant-in-Aid for Scientific Research (C) No.19K03863, Grant-in-Aid for Scientific Research (A) No.\,21H04469, and Grant-in-Aid for Transformative Research Areas (A) ``Extreme Universe''
No.\,21H05182 and No.\,21H05190.
The work of S.\,S. was supported by Grant-in-Aid for JSPS Fellows No.\,23KJ1533.

\appendix

\section{The detail of replica method}\label{appendix:replica method}
In this appendix, we provide an introductory review of the replica method in a fermionic theory and derive \eqref{eq: Replica integral}. In the following, we assume for simplicity that the fermionic theory posses the fermionic parity symmetry, namely the action $I[\psi, \overline{\psi}]$ satisfies
\begin{align}\label{eq: fermionic parity symmetry}
    I\,[\,\psi\,,\, \overline{\psi}\,]=I\,[\,-\psi\,,\, -\overline{\psi}\,]\ . 
\end{align}
To introduce the replica method, it is convenient to graphically represent the Euclidean path integrals. For instance, the transition amplitude from the initial state $\ket{\psi_{i}}$ to the final one $\ket{\psi_{f}}$ is drawn in the following way:
\begin{align}\label{eq:transition_amplitude}
    \bra{\psi_f} e^{-\beta H} \ket{\psi_i}
    &=
    \int^{\psi(x,\beta) = \psi_f}_{\psi(x,0) = \psi_i}
    \CD \psi \CD \overline{\psi}\ e^{-I[\psi, \overline{\psi}]}\nonumber\\
    &\equiv
    \begin{tikzpicture}[scale=0.8,baseline={([yshift=-.5ex]current bounding box.center)}]
        \tikzset{BENDTO/.style={color=black!50!white,very thin,looseness=0.5,bend left=60}}
        \coordinate(left_down)  at (0,0);
        \coordinate(right_down) at (4,0);
        \coordinate(left_up)  at (0,3);
        \coordinate(right_up) at (4,3);
        \fill[cyan!20] (left_down)--(right_down)--(right_up)--(left_up) --cycle;
        \draw[line width=1,dashed] (left_down)--(right_down);
        \draw[line width=1,dashed] (right_up)--(left_up);
        \draw[line width=1] (left_down)--(left_up);
        \draw[line width=1] (right_down)--(right_up);
        \draw ($(left_down)!0.5!(right_down)$) node[below] {$\psi_i$}; 
        \draw ($(left_up)!0.5!(right_up)$) node[above] {$\psi_f$}; 
        \draw (right_down) node[right=2mm] {$t_{\text{E}} = 0$};
        \draw (right_up) node[right=2mm] {$t_{\text{E}} = \beta $};
    \end{tikzpicture}
    .
\end{align}
In the second line, we perform the path integral over the blue region with the proper boundary conditions on the dashed lines.

We first derive the path integral representation of the reduced density matrix on an entangling region $V$ lying on the Euclidean time slice $t_{\text{E}}=0$. For later convenience, we define the fermionic fields $\psi_{V}$ and $\psi_{\overline{V}}$ as follows;
\begin{align}
\begin{aligned}
\psi_{V}\,(x)&\equiv \psi\,(x\in V \,,\, t_{\text{E}}=0) \\
    \psi_{\overline{V}}\,(x)&\equiv \psi\,(x\in \overline{V} \,,\, t_{\text{E}}=0)
\end{aligned}
\end{align}
and consider the state $\ket{\psi_V \oplus \psi_{\overline{V}}}$. Then, the path integral representations for $\braket{0|\psi_V \oplus \psi_{\overline{V}}}$ and $\braket{\psi_V \oplus \psi_{\overline{V}}|0}$ become as follows;
\begin{align}
    \braket{0|\psi_V \oplus \psi_{\overline{V}}}
    &=
    \frac{1}{\sqrt{\CN}} \int^{\psi(x,\infty)}_{\substack{\psi(x\in V,0) = \psi_V\\ \psi(x\in \overline{V},0) = \psi_{\overline{V}}}}
    \CD \psi \CD \overline{\psi}\ e^{-I[\psi, \overline{\psi}]}\ ,\nonumber\\
    &=
    \frac{1}{\sqrt{\CN}} \times
    \begin{tikzpicture}[scale=0.7, baseline={([yshift=-.5ex]current bounding box.center)}]
        \coordinate(left_down)  at (0,0);
        \coordinate(right_down) at (4,0);
        \coordinate(left_up)  at (0,3);
        \coordinate(right_up) at (4,3);
        \coordinate(u) at (1,0);
        \coordinate(v) at (3,0);
        \fill[cyan!20] (left_down)--(right_down)--(right_up)--(left_up) --cycle;
        \draw[line width=1] (left_down)--(left_up)--(right_up)--(right_down);
        \draw[line width=1,dashed] (left_down)--(u);
        \draw[line width=1,dashed,red!80] (u)--(v);
        \draw[line width=1,dashed] (v)--(right_down);
        \draw[red!80] ($(left_down)!0.5!(right_down)$) node[below] {$\psi_V$}; 
        \draw ($(left_down)!0.5!(u)$) node[below] {$\psi_{\overline{V}}$}; 
        \draw ($(v)!0.5!(right_down)$) node[below] {$\psi_{\overline{V}}$}; 
        \draw (right_down) node[right=2mm] {$t_{\text{E}} = 0$};
        \draw (right_up) node[right=2mm] {$t_{\text{E}} = \infty $};
        \fill[red!80](u) circle (2pt) (v) circle (2pt);
    \end{tikzpicture},\label{eq:<0|psi>_2}
\end{align}
\begin{align}
    \braket{\psi_V \oplus \psi_{\overline{V}}|0}
    &=
    \frac{1}{\sqrt{\CN}} \int_{\psi(x,-\infty)}^{\substack{\psi(x\in V,0) = \psi_V\\ \psi(x\in \overline{V},0) = \psi_{\overline{V}}}}
    \CD \psi \CD \overline{\psi}\ e^{-I[\psi, \overline{\psi}]}\ ,\nonumber\\
    &=
    \frac{1}{\sqrt{\CN}} \times
    \begin{tikzpicture}[scale=0.7, baseline={([yshift=-.5ex]current bounding box.center)}]
        \coordinate(left_down)  at (0,0);
        \coordinate(right_down) at (4,0);
        \coordinate(left_up)  at (0,3);
        \coordinate(right_up) at (4,3);
        \coordinate(u) at (1,3);
        \coordinate(v) at (3,3);
        \fill[cyan!20] (left_down)--(right_down)--(right_up)--(left_up) --cycle;
        \draw[line width=1] (left_up)--(left_down)--(right_down)--(right_up);
        \draw[line width=1,dashed] (left_up)--(u);
        \draw[line width=1,dashed,red!80] (u)--(v);
        \draw[line width=1,dashed] (v)--(right_up);
        \draw[red!80] ($(left_up)!0.5!(right_up)$) node[above] {$\psi_V$}; 
        \draw ($(left_up)!0.5!(u)$) node[above] {$\psi_{\overline{V}}$}; 
        \draw ($(v)!0.5!(right_up)$) node[above] {$\psi_{\overline{V}}$}; 
        \draw (right_down) node[right=2mm] {$t_{\text{E}} = -\infty$};
        \draw (right_up) node[right=2mm] {$t_{\text{E}} = 0 $};
        \fill[red!80](u) circle (2pt) (v) circle (2pt);
    \end{tikzpicture}\label{eq:<psi|0>_2}\ , 
\end{align}
where $\CN$ is some normalization factor and will be determined soon. In the above expression, we schematically illustrate the bunch of intervals by red dashed line. We next derive the path integral representation of the reduced density matrix $\rho_{V}$. By using the aforementioned results, the matrix element $\bra{\psi_V} \rho_V \ket{\psi'_{V}}$ of the reduced density matrix can be formulated as follows;
\begin{align}
    &\bra{\psi_V} \rho_V \ket{\psi'_{V}}\nonumber  \\ 
    &\equiv
    \int \CD \psi_{\overline{V}}  \CD \overline{\psi_{\overline{V}}} 
    \braket{\psi_V \oplus (-\psi_{\overline{V}})|0} \braket{0|\psi'_V \oplus \psi_{\overline{V}}}\ ,\nonumber\\
    &=
    \frac{1}{\CN} \int \CD \psi_{\overline{V}} \CD \overline{\psi_{\overline{V}}}
    \begin{tikzpicture}[scale=0.8, baseline={([yshift=-.5ex]current bounding box.center)}]
        \coordinate(left_down)  at (0,0);
        \coordinate(right_down) at (4,0);
        \coordinate(left_up)  at (0,1.5);
        \coordinate(right_up) at (4,1.5);
        \coordinate(u) at (1,1.5);
        \coordinate(v) at (3,1.5);
        \coordinate(s) at (0,2.5); 
        \coordinate(left_down_s)  at ($(left_down)+(s)$);
        \coordinate(right_down_s) at ($(right_down)+(s)$);
        \coordinate(left_up_s)  at ($(left_up)+(s)$);
        \coordinate(right_up_s) at ($(right_up)+(s)$);
        \coordinate(u_s) at ($(u)+(s)-(left_up)$);
        \coordinate(v_s) at ($(v)+(s)-(left_up)$);
        \fill[cyan!20] (left_down)--(right_down)--(right_up)--(left_up) --cycle;
        \fill[cyan!20] (left_down_s)--(right_down_s)--(right_up_s)--(left_up_s) --cycle;
        \draw[line width=1] (left_up)--(left_down)--(right_down)--(right_up);
        \draw[line width=1,dashed] (left_up)--(u);
        \draw[line width=1,dashed,red!80] (u)--(v);
        \draw[line width=1,dashed] (v)--(right_up);
        \draw[line width=1] (left_down_s)--(left_up_s)--(right_up_s)--(right_down_s);
        \draw[line width=1,dashed] (left_down_s)--(u_s);
        \draw[line width=1,dashed,red!80] (u_s)--(v_s);
        \draw[line width=1,dashed] (v_s)--(right_down_s);
        \draw[red!80] ($(left_up)!0.5!(right_up)$) node[below] {$\psi_V$}; 
        \draw ($(left_up)!0.5!(u)$) node[below] {$-\psi_{\overline{V}}$}; 
        \draw ($(v)!0.5!(right_up)$) node[below] {$-\psi_{\overline{V}}$}; 
        \draw (right_down) node[right=2mm] {$t_{\text{E}} = -\infty$};
        \draw (right_up) node[right=2mm] {$t_{\text{E}} = -0 $};
        \draw[red!80] ($(left_down_s)!0.5!(right_down_s)$) node[above] {$\psi'_V$}; 
        \draw ($(left_down_s)!0.5!(u_s)$) node[above] {$\psi_{\overline{V}}$}; 
        \draw ($(v_s)!0.5!(right_down_s)$) node[above] {$\psi_{\overline{V}}$}; 
        \draw (right_down_s) node[right=2mm] {$t_{\text{E}} = +0$};
        \draw (right_up_s) node[right=2mm] {$t_{\text{E}} = \infty $};
        \fill[red!80](u) circle (2pt) (v) circle (2pt);
        \fill[red!80](u_s) circle (2pt) (v_s) circle (2pt);
    \end{tikzpicture},\nonumber\\ \nonumber\\
    &=
     \frac{1}{\CN} 
     \textcolor{black}{
     \int \CD \psi_{\overline{V}} \CD \overline{\psi_{\overline{V}}} }
     \begin{tikzpicture}[scale=0.8, baseline={([yshift=-.5ex]current bounding box.center)}]
         \coordinate(left_down)  at (0,0);
     \coordinate(right_down) at (4,0);
     \coordinate(left_up)  at (0,1.5);
         \coordinate(right_up) at (4,1.5);
         \coordinate(u) at (1,1.5);
         \coordinate(v) at (3,1.5);
         \coordinate(s) at (0,2.5); 
         \coordinate(left_down_s)  at ($(left_down)+(s)$);
         \coordinate(right_down_s) at ($(right_down)+(s)$);
         \coordinate(left_up_s)  at ($(left_up)+(s)$);
         \coordinate(right_up_s) at ($(right_up)+(s)$);
         \coordinate(u_s) at ($(u)+(s)-(left_up)$);
         \coordinate(v_s) at ($(v)+(s)-(left_up)$);
         \fill[cyan!20] (left_down)--(right_down)--(right_up)--(left_up) --cycle;
         \fill[cyan!20] (left_down_s)--(right_down_s)--(right_up_s)--(left_up_s) --cycle;
         \draw[line width=1] (left_up)--(left_down)--(right_down)--(right_up);
         \draw[line width=1,dashed] (left_up)--(u);
        \draw[line width=1,dashed,red!80] (u)--(v);
         \draw[line width=1,dashed] (v)--(right_up);
         \draw[line width=1] (left_down_s)--(left_up_s)--(right_up_s)--(right_down_s);
         \draw[line width=1,dashed] (left_down_s)--(u_s);
         \draw[line width=1,dashed,red!80] (u_s)--(v_s);
         \draw[line width=1,dashed] (v_s)--(right_down_s);
         \draw[red!80] ($(left_up)!0.5!(right_up)$) node[below] {$-\psi_V$}; 
         \draw ($(left_up)!0.5!(u)$) node[below] {\textcolor{black}{$\psi_{\overline{V}}$}}; 
         \draw ($(v)!0.5!(right_up)$) node[below] {\textcolor{black}{$\psi_{\overline{V}}$}}; 
         \draw (right_up) node[right=2mm] {$t_{\text{E}} = -0 $};
         \draw[red!80] ($(left_down_s)!0.5!(right_down_s)$) node[above] {$\psi'_V$}; 
         \draw ($(left_down_s)!0.5!(u_s)$) node[above] {\textcolor{black}{$\psi_{\overline{V}}$}}; 
         \draw ($(v_s)!0.5!(right_down_s)$) node[above] {\textcolor{black}{$\psi_{\overline{V}}$}}; 
         \draw (right_down_s) node[right=2mm] {$t_{\text{E}} = +0$};
         \fill[red!80](u) circle (2pt) (v) circle (2pt);
         \fill[red!80](u_s) circle (2pt) (v_s) circle (2pt);
         \draw ($(left_up)!0.5!(u_s)$) node {\textcolor{teal}{\scalebox{1}{$\updownarrow $}}};
         \draw ($(right_up)!0.5!(v_s)$) node {\textcolor{teal}{\scalebox{1}{$\updownarrow $}}};
         \draw ($(u)!0.5!(v_s)$) node {\textcolor{teal}{gluing}};
     \end{tikzpicture},\nonumber\\ \nonumber\\
     &=
    \frac{1}{\CN} \times 
    \begin{tikzpicture}[scale=0.8, baseline={([yshift=-.5ex]current bounding box.center)}]
        \coordinate(left_down)  at (0,0);
        \coordinate(right_down) at (4,0);
        \coordinate(left_up)  at (0,3);
        \coordinate(right_up) at (4,3);
        \coordinate(epsilon) at (0,0.1);
        \coordinate(u_up) at ($(1,3/2)+(epsilon)$);
        \coordinate(v_up) at ($(3,3/2)+(epsilon)$);
        \coordinate(u_down) at ($(1,3/2)-(epsilon)$);
        \coordinate(v_down) at ($(3,3/2)-(epsilon)$);
        \fill[cyan!20] (left_down)--(right_down)--(right_up)--(left_up) --cycle;
        \fill[white] (u_down)--(v_down)--(v_up)--(u_up) --cycle;
        \draw[line width=1] (left_up)--(left_down)--(right_down)--(right_up)--cycle;
        \draw[line width=1,dashed,red!80] (u_down)--(v_down)--(v_up)--(u_up) --cycle;
        \draw[red!80] ($(u_up)!0.5!(v_up)$) node[above] {$\psi'_V$}; 
        \draw[red!80] ($(u_down)!0.5!(v_down)$) node[below] {$-\psi_V$}; 
        \draw ($(right_down)!0.5!(right_up) + 2*(epsilon)$) node[right=2mm] {$t_{\text{E}} = +0$};
        \draw ($(right_down)!0.5!(right_up) - 2*(epsilon)$) node[right=2mm] {$t_{\text{E}} = -0$};
    \end{tikzpicture}\ ,\nonumber\\ \nonumber\\
    &=
    \frac{1}{\CN} \int_{\substack{
        \psi(x\in V,+0) = \psi'_V\\ 
        \psi(x\in V,-0) = -\psi_{V}}}
        \CD \psi \CD \overline{\psi} \ e^{-I[\psi, \overline{\psi}]}\ .
    \label{eq:reduced_density_matrix_path_integral}
\end{align}
Note that the minus sign in the boundary condition in the second line comes from the fermionic trace $\Tr_{\overline {V}}\left[\mathcal{O}\right] \equiv \int \CD \psi_{\overline{V}} \CD \overline{\psi_{\overline{V}}}\, \braket{-\psi_{\overline{V}} |\mathcal{O}|\psi_{\overline{V}}}$.
When going from the third line to the forth, we use the fermion parity symmetry described in \eqref{eq: fermionic parity symmetry}. Eventually, after gluing the integration variable $\psi_{\overline{V}}$ outside the entangling region $V$, we arrive at the path integral expression for the reduced density matrix. The reduced density matrix must satisfy $1 = \Tr_V \left[\rho_V\right] = \int \CD \psi_{V}  \CD \overline{\psi_{V}} \bra{-\psi_V} \rho_V \ket{\psi_{V}}$, from which the normalization factor $\CN$ must be equal to the partition function $Z_{1}$ without an entangling region:
\begin{align}
    \CN=Z_{1}\ . 
\end{align}

Let us derive the formula \eqref{eq: Replica integral} for $n=2$. The generalization to the case with $n\geq 3$ can be straightforwardly performed. From the path integral representation of the reduced density matrix \eqref{eq:reduced_density_matrix_path_integral}, we obtain;
\begin{align}
    \Tr_V \left[\rho^2_V\right]
    &=
    \int \CD \psi_V \CD \overline{\psi_V}  \CD \psi'_V  \CD \overline{\psi'_V} \nonumber\\
    &\hspace{10mm} \times
    \bra{-\psi_V} \rho_V \ket{\psi'_{V}} \bra{\psi_V'} \rho_V \ket{\psi_{V}} , \nonumber\\
    &=
    \frac{1}{Z^2_1} \textcolor{black}{\int \CD \psi_V \CD \overline{\psi_V}}  \textcolor{black}{\CD \psi'_V \CD \overline{\psi_V'}} \nonumber\\
    &\hspace{10mm}
    \begin{tikzpicture}[scale=0.7, baseline={([yshift=-.5ex]current bounding box.center)}]
        \coordinate(left_down)  at (0,0);
        \coordinate(right_down) at (4,0);
        \coordinate(left_up)  at (0,3);
        \coordinate(right_up) at (4,3);
        \coordinate(epsilon) at (0,0.1);
        \coordinate(u_up) at ($(1,3/2)+(epsilon)$);
        \coordinate(v_up) at ($(3,3/2)+(epsilon)$);
        \coordinate(u_down) at ($(1,3/2)-(epsilon)$);
        \coordinate(v_down) at ($(3,3/2)-(epsilon)$);
        \fill[cyan!20] (left_down)--(right_down)--(right_up)--(left_up) --cycle;
        \fill[white] (u_down)--(v_down)--(v_up)--(u_up) --cycle;
        \draw[line width=1] (left_up)--(left_down)--(right_down)--(right_up)--cycle;
        \draw[line width=1,dashed,red!80] (u_down)--(v_down)--(v_up)--(u_up) --cycle;
        \draw[black] ($(u_up)!0.5!(v_up)$) node[above] {$\psi'_V$}; 
        \draw[black] ($(u_down)!0.5!(v_down)$) node[below] {$\psi_V$}; 
        \draw (left_down) node[above right=1mm] {$\psi_1$}; 
        \coordinate(s)  at (5,0);
        \coordinate(left_down2)  at ($(left_down)+(s)$);
        \coordinate(right_down2) at ($(right_down)+(s)$);
        \coordinate(left_up2)  at ($(left_up)+(s)$);
        \coordinate(right_up2) at ($(right_up)+(s)$);
        \coordinate(u_up2) at ($(u_up)+(s)$);
        \coordinate(v_up2) at ($(v_up)+(s)$);
        \coordinate(u_down2) at ($(u_down)+(s)$);
        \coordinate(v_down2) at ($(v_down)+(s)$);
        \fill[cyan!20] (left_down2)--(right_down2)--(right_up2)--(left_up2) --cycle;
        \fill[white] (u_down2)--(v_down2)--(v_up2)--(u_up2) --cycle;
        \draw[line width=1] (left_up2)--(left_down2)--(right_down2)--(right_up2)--cycle;
        \draw[line width=1,dashed,red!80] (u_down2)--(v_down2)--(v_up2)--(u_up2) --cycle;
        \draw[black] ($(u_up2)!0.5!(v_up2)$) node[above] {$\psi_V$}; 
        \draw[black] ($(u_down2)!0.5!(v_down2)$) node[below] {$-\psi'_V$};
        \draw (left_down2) node[above right=1mm] {$\psi_2$}; 
    \end{tikzpicture}\ ,\nonumber\\ \nonumber \\
    &=
    \frac{1}{Z^2_1} \textcolor{black}{\int \CD \psi_V \CD \overline{\psi_V}}  \textcolor{black}{\CD \psi'_V \CD \overline{\psi'_V}} \nonumber\\
    &\hspace{10mm}
    \begin{tikzpicture}[scale=0.7, baseline={([yshift=-.5ex]current bounding box.center)}]
        \coordinate(left_down)  at (0,0);
        \coordinate(right_down) at (4,0);
        \coordinate(left_up)  at (0,3);
        \coordinate(right_up) at (4,3);
        \coordinate(epsilon) at (0,0.1);
        \coordinate(u_up) at ($(1,3/2)+(epsilon)$);
        \coordinate(v_up) at ($(3,3/2)+(epsilon)$);
        \coordinate(u_down) at ($(1,3/2)-(epsilon)$);
        \coordinate(v_down) at ($(3,3/2)-(epsilon)$);
        \fill[cyan!20] (left_down)--(right_down)--(right_up)--(left_up) --cycle;
        \fill[white] (u_down)--(v_down)--(v_up)--(u_up) --cycle;
        \draw[line width=1] (left_up)--(left_down)--(right_down)--(right_up)--cycle;
        \draw[line width=1,dashed,red!80] (u_down)--(v_down)--(v_up)--(u_up) --cycle;
        \draw[black] ($(u_up)!0.5!(v_up)$) node[above] {$\psi'_V$}; 
        \draw[black] ($(u_down)!0.5!(v_down)$) node[below] {$\psi_V$}; 
        \draw (left_down) node[above right=1mm] {$\psi_1$}; 
        \coordinate(s)  at (5,0);
        \coordinate(left_down2)  at ($(left_down)+(s)$);
        \coordinate(right_down2) at ($(right_down)+(s)$);
        \coordinate(left_up2)  at ($(left_up)+(s)$);
        \coordinate(right_up2) at ($(right_up)+(s)$);
        \coordinate(u_up2) at ($(u_up)+(s)$);
        \coordinate(v_up2) at ($(v_up)+(s)$);
        \coordinate(u_down2) at ($(u_down)+(s)$);
        \coordinate(v_down2) at ($(v_down)+(s)$);
        \fill[cyan!20] (left_down2)--(right_down2)--(right_up2)--(left_up2) --cycle;
        \fill[white] (u_down2)--(v_down2)--(v_up2)--(u_up2) --cycle;
        \draw[line width=1] (left_up2)--(left_down2)--(right_down2)--(right_up2)--cycle;
        \draw[line width=1,dashed,red!80] (u_down2)--(v_down2)--(v_up2)--(u_up2) --cycle;
        \draw[black] ($(u_up2)!0.5!(v_up2)$) node[above] {$-\psi_V$}; 
        \draw[black] ($(u_down2)!0.5!(v_down2)$) node[below] {$\psi'_V$}; 
        \draw (left_down2) node[above right=1mm] {$\psi_2$}; 
        \draw[orange,line width=1] ($(u_up)!0.5!(v_up)+(0.5,0.5)$) to [out=30, in=-150] ($(u_down2)!0.5!(v_down2)+(-0.5,-0.5)$);
        \draw[teal,line width=1] ($(u_up2)!0.5!(v_up2)+(0.5,0.5)$) to [out=30, in=90] ($(right_down2)!0.5!(right_up2)+(1,0)$) to [out=-90, in=-90] ($(u_down)!0.5!(v_down)+(0,-0.8)$);
    \end{tikzpicture}\ ,\nonumber\\
    &=
    \frac{1}{Z^2_1} \times
    \begin{tikzpicture}[scale=0.5, baseline={([yshift=-.5ex]current bounding box.center)}]
        \coordinate(left_down)  at (0,0);
        \coordinate(right_down) at (10,0);
        \coordinate(left_up)  at (2,2);
        \coordinate(right_up) at (12,2);
        \coordinate(epsilon) at (0.1,0.1);
        \coordinate(u_up) at ($(4,1)+(epsilon)$);
        \coordinate(v_up) at ($(8,1)+(epsilon)$);
        \coordinate(u_down) at ($(4,1)-(epsilon)$);
        \coordinate(v_down) at ($(8,1)-(epsilon)$);
        \fill[cyan!20]
        (left_down)--(right_down)--(right_up)--(left_up)--cycle;
        \fill[white]
        (u_down)--(v_down)--(v_up)--(u_up)--cycle;
        \draw[line width=1]
        (left_down)--(right_down)--(right_up)--(left_up)--cycle;
        \draw[line width=1,red!80]
        (u_down)--(v_down)--(v_up)--(u_up)--cycle;
        \draw ($(left_down)+(1.5,0.5)$) node {\scalebox{1}{\rotatebox{-10}{$\psi_1$}}};
        \coordinate(s)  at (0,2.5); 
        \coordinate(left_down2)  at ($(left_down)+(s)$);
        \coordinate(right_down2) at ($(right_down)+(s)$);
        \coordinate(left_up2)  at ($(left_up)+(s)$);
        \coordinate(right_up2) at ($(right_up)+(s)$);
        \coordinate(u_up2) at ($(u_up)+(s)$);
        \coordinate(v_up2) at ($(v_up)+(s)$);
        \coordinate(u_down2) at ($(u_down)+(s)$);
        \coordinate(v_down2) at ($(v_down)+(s)$);
        \fill[cyan!20]
        (left_down2)--(right_down2)--(right_up2)--(left_up2)--cycle;
        \fill[white]
        (u_down2)--(v_down2)--(v_up2)--(u_up2)--cycle;
        \draw[line width=1]
        (left_down2)--(right_down2)--(right_up2)--(left_up2)--cycle;
        \draw[line width=1,red!80]
        (u_down2)--(v_down2)--(v_up2)--(u_up2)--cycle;
        \draw ($(left_down2)+(1.5,0.5)$) node {\scalebox{1}{\rotatebox{-10}{$\psi_2$}}};
        \coordinate(ua) at ($(u_up2)+0.5*(s)-0.5*(epsilon)$);
        \coordinate(va) at ($(v_up2)+0.5*(s)-0.5*(epsilon)$);
        \coordinate(ua_top) at ($(u_up2)+0.7*(s)+0.1*(v_up)-0.1*(u_up)$);
        \coordinate(va_top) at ($(v_up2)+0.7*(s)+0.1*(v_up)-0.1*(u_up)$);
        \coordinate(va_right) at ($(v_up2)+0.5*(s)-0.5*(epsilon)+0.2*(v_up)-0.2*(u_up)$);
        \coordinate(u0) at ($(u_down) - 0.2*(s) + (0.1,-0.1)$);
        \coordinate(v0) at ($(v_down) - 0.2*(s) + (0.1,-0.1)$);
        \coordinate(v0_right) at ($(v_up)-0.2*(s) -0.5*(epsilon)+0.2*(v_up)-0.2*(u_up)+ (0,0.5)$);
        \fill[lightgray!80] 
        (u_up)--(v_up)--($(v_up)!0.59!(v_down2)$)--($(u_up)!0.59!(u_down2)$)--cycle;
        \fill[lightgray!20]
        ($(v_up)!0.63!(v_down2)$)--
        ($(u_up)!0.63!(u_down2)$)--
        ($(u_down2) -(0,0.05)$)--
        ($(v_down2)-(0,0.05) $)--cycle;
        \fill[lightgray!80] 
        (u_up2) to (v_up2) to (va) to [out=95, in=180] (va_top) to (ua_top) to [out=180,in=95] (ua) to (u_up2)--cycle;
        \draw[line width=1,lightgray!80]
        (va_top) to [out=0,in=90] (va_right) to ($(va_right)!0.07!(v0_right)$);
        \draw[line width=1,lightgray!20,dashed]
        ($(va_right)!0.1!(v0_right)$)--($(va_right)!0.55!(v0_right)$);
        \draw[line width=1,lightgray!80]
        ($(va_right)!0.63!(v0_right)$)--($(va_right)!0.73!(v0_right)$);
        \draw[line width=1,lightgray!20,dashed]
        ($(va_right)!0.8!(v0_right)$) to (v0_right) to [out=-90,in=0] (v0);
        \fill[lightgray!20]
        ($(u_down)-(0,0.05)$)--($(v_down)-(0,0.05)$)--(v0)--(u0)--cycle;
        \draw[black] ($(u_up)!0.5!(v_up)$) node[above] {\scalebox{0.8}{$\psi'_V$}}; 
        \draw[black] ($(u_down)!0.5!(v_down)$) node[below] {\scalebox{0.8}{$\psi_V$}}; 
        \draw[black] ($(u_up2)!0.5!(v_up2)$) node[above] {\scalebox{0.8}{$-\psi_V$}}; 
        \draw[black] ($(u_down2)!0.5!(v_down2)$) node[below] {\scalebox{0.8}{$\psi'_V$}}; 
  \end{tikzpicture}\ ,\nonumber \\ \nonumber \\
  &=
    \frac{1}{Z^2_1} \int_{\text{B.C.}}
    \CD \psi_1 \CD \overline{\psi_1} \CD \psi_2 \CD \overline{\psi_2} \ e^{-I[\psi_1,\overline{\psi_1}]-I[\psi_2,\overline{\psi_2} ]}\ ,
  \label{eq:Tr_V_rho_V^2}
\end{align}
where $\psi_i, (i=1,2)$ represents the fermion field on the $i$-th sheet.
When going from the second line to the third one, we change the path integral variable on the second sheet as $\psi_2 \to -\psi_2$. The boundary condition (denoted by B.C.) in the last line means
\begin{align}
\text{B.C.}\  \ \left\{
\begin{aligned}
    \psi_1 (x\in V,+0) &= \psi_2 (x\in V,-0)\ ,\\ 
    \psi_2 (x\in V,+0) &= -\psi_1 (x\in V,-0)\ .
\end{aligned}
\right. 
\end{align}
This completes the proof for the formula \eqref{eq: Replica integral} in the case of $n=2$.

\section{Formulas on theta function and Dedekind eta function}\label{sec: Theta function and Eta function identities}
In this appendix, we list some formulas relating the theta function and the Dedekind eta function which are widely used throughout this paper.

\paragraph{Jacobi identity.}
\begin{align}\label{eq: theta_func_jacobi_id}
    \vartheta_{3}(\tau)^4 - \vartheta_{2}(\tau)^4 - \vartheta_{4}(\tau)^4 =0 \ .
\end{align}

\paragraph{Eta and theta relation.}
\begin{align}\label{eq: eta_func_theta_func_relation}
    2 \eta(\tau)^3 = \vartheta_{2}(\tau)\vartheta_{3}(\tau)\vartheta_{4}(\tau)\ .
\end{align}

\paragraph{Doubling identities.}
\begin{align}
    \vartheta_{2}(2 \tau) &= \left( \frac{\vartheta_{3}(\tau)^2 - \vartheta_{4}(\tau)^2}{2} \right)^\frac{1}{2}\ ,\label{theta_2_double_angle_formula}\\
    \vartheta_{3}(2 \tau) &= \left( \frac{\vartheta_{3}(\tau)^2 + \vartheta_{4}(\tau)^2}{2} \right)^\frac{1}{2}\ ,\label{theta_3_double_angle_formula}\\
    \vartheta_{4}(2 \tau) &= \left( \vartheta_{3}(\tau)\   \vartheta_{4}(\tau) \right)^\frac{1}{2}\ .\label{theta_4_double_angle_formula}
\end{align}
\paragraph{Modular properties.}
\begin{align}
    \vartheta_{2} \left( - \frac{1}{\tau} \right) = \sqrt{- \i \,\tau}\  \vartheta_{4} (\tau) \ ,\label{modular_S_trsf_theta2}\\
    \vartheta_{3} \left( - \frac{1}{\tau} \right) = \sqrt{- \i \,\tau}\  \vartheta_{3} (\tau) \ ,\label{modular_S_trsf_theta3}\\
    \vartheta_{4} \left( - \frac{1}{\tau} \right) = \sqrt{- \i \,\tau}\  \vartheta_{2} (\tau) \ ,\label{modular_S_trsf_theta4}\\
    \eta \left( - \frac{1}{\tau} \right) = \sqrt{- \i \,\tau}      \  \eta (\tau) \ .\label{modular_S_trsf_eta}
\end{align}

\paragraph{Asymptotic behaviors.} 
\begin{align}
\begin{aligned}
        \vartheta_{2} (\i \,\ell ) 
    &\sim
    2 e^{- \frac{\pi \ell}{4}}\ , \\
    \vartheta_{3} (\i \,\ell ) 
    &\sim
    1 + 2 e^{-\pi \ell} \ , \qquad \text{as} \quad \ell \sim \infty\ ,\label{limit_theta3_l_infty}\\
    \vartheta_{4} (\i \,\ell ) 
    &\sim
    1 - 2 e^{-\pi \ell} \ . \ \  
\end{aligned}
\end{align}
\begin{align}
\begin{aligned}
        \vartheta_{2}(\i \,\ell) 
    &\sim   
    \ell^{-1/2} \ , \\ 
    \vartheta_{3}(\i \,\ell) 
    &\sim 
    \ell^{-1/2} \ , \qquad \qquad  \text{as} \quad \ell \sim 0 \ ,\label{limit_theta3_l_+0} \\
    \vartheta_{4}(\i \,\ell) 
    &\sim
    2\, \ell^{-1/2}\, e^{-\frac{\pi}{4 \ell}} \ . 
\end{aligned}
\end{align}

\bibliography{EE_Ferm}

\end{document}